%% file: arXiv.tex
\documentclass[journal]{IEEEtai}
\input{premable}

\usepackage{xspace}
\newcommand{\modelname}{DiffPlace\xspace}

\usepackage[colorlinks,urlcolor=blue,linkcolor=blue,citecolor=blue]{hyperref}

\usepackage{color,array}

\usepackage{graphicx}

\begin{document}

\title{\modelname: A Conditional Diffusion Framework for Simultaneous VLSI Placement Beyond Sequential Paradigms}

\author{Kien Le Trung
\thanks{Corresponding Author: Truong-Son Hy}
\thanks{K. Le Trung is with The University of Alabama at Birmingham, United States (email: kien.lt203474@gmail.com)}
\and Truong-Son Hy
\thanks{T. S. Hy is with The University of Alabama at Birmingham, United States (email: thy@uab.edu)}}

\markboth{Journal of IEEE Transactions on Artificial Intelligence, Vol. 00, No. 0, Month 2026}
{Le \& Hy: \modelname: A Conditional Diffusion Framework for Simultaneous VLSI Placement Beyond Sequential Paradigms}

\maketitle

\begin{abstract}
\label{sec:abstract}
Chip placement, a critical step in the VLSI physical design flow, directly impacts performance, power, and routability. Traditional chip placement methods, relying on analytical optimization or sequential reinforcement learning (RL), face significant challenges in modern VLSI design, including the inability to consistently satisfy hard placement constraints and the requirement for computationally expensive online training for each new circuit design. Furthermore, existing sequential decision-making paradigms often suffer from compounding errors and extreme wirelength minimization that aggressively compresses modules into dense clusters, leading to severe routing congestion hotspots and failures in downstream design stages. To address these limitations, we introduce \modelname, a framework that reformulates chip placement as a conditional denoising diffusion process, enabling transferable policies that generalize to unseen netlists without extensive retraining. Unlike sequential paradigms, \modelname simultaneously optimizes all macro positions utilizing a neural backbone equipped with vector-wise message passing to capture geometric dependencies. By prioritizing a more balanced spatial distribution of macros, our framework adopts a routability-first perspective to effectively prevent routing hotspots while maintaining competitive wirelength. To effectively handle the multi-objective nature of placement, we propose a decoupled guidance mechanism: global objectives are optimized via energy-based conditioning, while local physical constraints are actively mitigated through explicit manifold gradient injection during the reverse sampling process. Extensive experiments demonstrate that \modelname achieves competitive placement quality while offering superior generalization efficiency compared to state-of-the-art learning-based baselines. 

We provide a demonstration implementation of the proposed method at: \url{https://github.com/HySonLab/DiffPlace}.
\end{abstract}

\begin{IEEEImpStatement}
This paper introduces \modelname, a fundamentally new paradigm for VLSI chip placement that shifts the field from sequential decision-making to parallel generative modeling. Unlike prior reinforcement learning (RL) and analytical methods that either suffer from compounding errors or lack cross-design knowledge transfer, DiffPlace formulates placement as a conditional diffusion process, enabling simultaneous optimization of all module positions and eliminating sequential bottlenecks. 
The key leap in this work lies in unifying generative AI with physical design constraints. Through energy-based conditioning and decoupled guidance, DiffPlace explicitly separates global objectives (e.g., wirelength) from local constraints (e.g., overlap and congestion), overcoming instability issues that limit existing multi-objective optimization frameworks . Additionally, the proposed vector-wise message passing architecture introduces geometric inductive bias, allowing the model to capture translation-invariant spatial relationships critical for layout reasoning, an aspect underexplored in prior GNN-based placers.
Empirically, \modelname demonstrates strong zero-shot generalization to unseen industrial-scale designs, achieving competitive or superior placement quality while drastically reducing the need for costly per-design retraining. This represents a significant advancement over RL-based approaches, which require extensive online optimization for each new circuit.
Beyond performance gains, this work establishes a new research direction: leveraging diffusion models as scalable, transferable solvers for combinatorial optimization in EDA. The framework's ability to integrate learning, physics-based constraints, and generative modeling opens the door to next-generation AI-driven design automation systems with improved efficiency, robustness, and adaptability.
\end{IEEEImpStatement}


\input{sections/introduction} 
\input{sections/relatedworked}
\input{sections/preliminaries}
\input{sections/methodology}

\input{sections/experiments}

\input{sections/conclusion}
\bibliographystyle{IEEEtran}
\bibliography{references}


\clearpage

\input{sections/appendix}

\end{document}

%% file: premable.tex
\usepackage{amsmath,amsfonts}
\usepackage{amsmath}
\usepackage{array}
\usepackage{multirow}
\usepackage{tikz}
\usepackage[caption=false,font=normalsize,labelfont=sf,textfont=sf]{subfig}
\usepackage{textcomp}
\usepackage{stfloats}
\usepackage{url}
\usepackage{verbatim}
\usepackage{graphicx}
\usepackage{subcaption}
\usepackage{subfig}
\usepackage{cite}
\usepackage{algorithm}
\usepackage{algpseudocode}
\usepackage{tikz}
\usepackage{pgfplots}
\usepackage{booktabs}
\usepackage{amsmath}     
\pgfplotsset{compat=1.18}
\usepackage{pgfplots}
\usepackage{tikz}
\usepackage{subfig}
\pgfplotsset{compat=1.18}
\usetikzlibrary{intersections}
\usepgfplotslibrary{fillbetween}

\def\BibTeX{{\rm B\kern-.05em{\sc i\kern-.025em b}\kern-.08em
T\kern-.1667em\lower.7ex\hbox{E}\kern-.125emX}}
\usepackage{balance}

%% file: sections/introduction.tex
\section{Introduction}
\label{sec:introduction}

Modern integrated circuit (IC) design faces formidable challenges as semiconductor technology advances into the angstrom era, characterized by billions of transistors and stringent design rules. Within the electronic design automation (EDA) flow, \textit{chip placement}---the process of determining physical coordinates for circuit modules on a 2D canvas---remains a critical bottleneck. This stage fundamentally determines the lower bounds of power, performance, and area (PPA), as well as the feasibility of downstream routing. Yet, achieving optimal placement remains a profoundly challenging NP-hard problem. Circuit components must be arranged to minimize wirelength and timing delays while simultaneously adhering to complex, non-differentiable constraints such as density targets and routing congestion.

Traditionally, analytical solvers have served as the foundation of physical design. While these methods provide a global view of the optimization landscape, they often struggle to strictly enforce discrete constraints without disrupting the convergence of the global objective. In contrast, the emergence of reinforcement learning (RL) has catalyzed a shift toward data-driven placement. Frameworks such as GraphPlace \cite{DBLP:journals/corr/abs-2004-10746} and MaskPlace \cite{DBLP:conf/nips/LaiM022} have reconceptualized placement as a sequential decision process, demonstrating the ability to surpass human experts in specific scenarios. However, despite their promise, existing learning-based approaches are limited by several fundamental constraints:

First, the dominant paradigm relies on sequential decision-making. RL agents typically place macros one by one, where early suboptimal decisions irreversibly constrain the solution space for subsequent modules. This greedy nature introduces compounding errors that are computationally expensive to correct in later stages.

Second, current methods exhibit poor generalization efficiency. Due to the reliance on design-specific state spaces, RL agents necessitate costly online training from scratch for each new netlist. Even recent transfer learning attempts like ChiPFormer \cite{DBLP:conf/icml/LaiLTWH023} remain bound to the sequential framework, where adapting to unseen topologies necessitates extensive interactions or iterative policy updates to overcome compounding decision errors.

Third, the neural backbones used in prior works lack of sufficient geometric inductive biases.  Graph Neural Networks (GNNs) used in placement typically aggregate scalar features, failing to explicitly capture the relative vector relationships and translation invariance that are critical for understanding physical layout structures.

Fourth, optimizing for conflicting objectives remains a hurdle. Driven primarily by Half-Perimeter Wirelength (HPWL) minimization, existing models produce dense clusters that, while optimizing wirelength, trigger severe routing congestion and design rule violations (DRC) \cite{DBLP:conf/dac/LinC14, DBLP:conf/dac/XuLTLSSP19, DBLP:conf/dac/HeHCKLCY13}. Coupling these competing objectives into a single scalar reward function often leads to optimization instability.

To address these challenges, we introduce \modelname, a unified conditional diffusion framework designed to capture the joint distribution of all module positions for simultaneous, constraint-aware placement.

First, we transcend the sequential paradigm by formulating placement as a simultaneous denoising process. Unlike RL agents, \modelname gradually refines the positions of all macros in parallel, transforming random noise into a valid layout. This holistic approach eliminates compounding errors and allows the model to explore the global solution space more effectively.

Second, we enable efficient few-shot generalization through energy-based conditioning. By training on a diverse set of synthetic and real netlists, \modelname learns a transferable policy that can generate high-quality placements for unseen designs in a single inference pass, or adapt with minimal fine-tuning.

Third, to capture complex geometric dependencies, we propose a Vector-wise Message Passing architecture. This backbone explicitly encodes relative position vectors, ensuring the model learns physically meaningful, translation-invariant spatial relationships.

Fourth, we propose a Decoupled Guidance Mechanism to resolve objective conflicts. We optimize global objectives (HPWL) via implicit energy conditioning, while actively mitigating local physical constraints (overlap, congestion) through explicit manifold gradient injection during the reverse sampling process.

\begin{figure*}[t]
    \centering
    \begin{tabular}{@{}cccccc@{}}
        \includegraphics[width=0.15\textwidth]{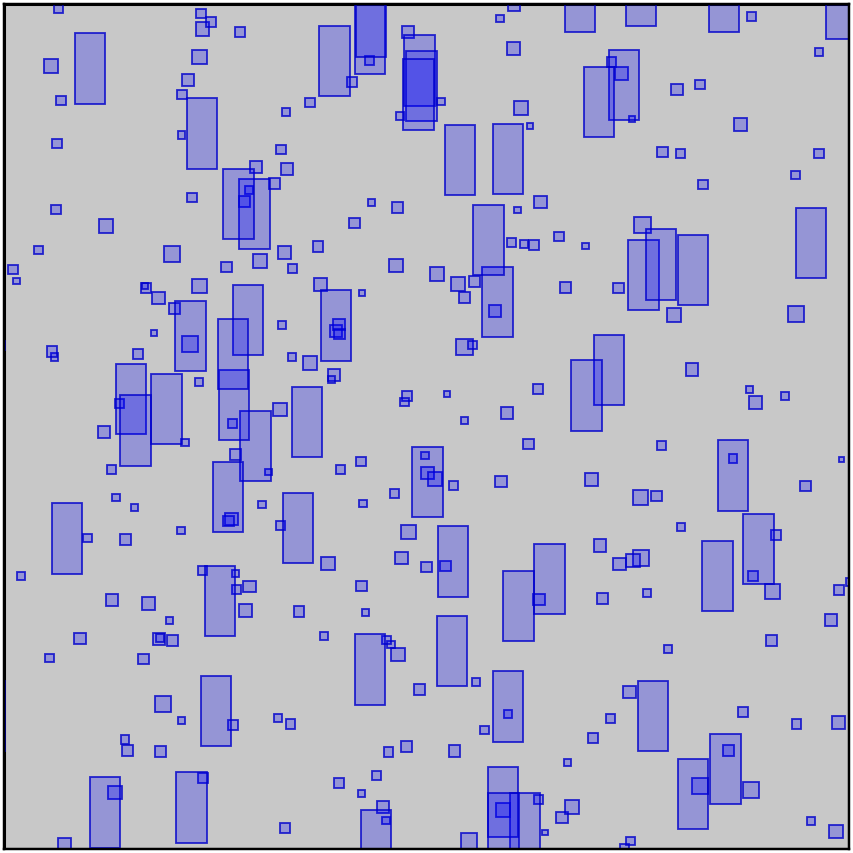} &
        \includegraphics[width=0.15\textwidth]{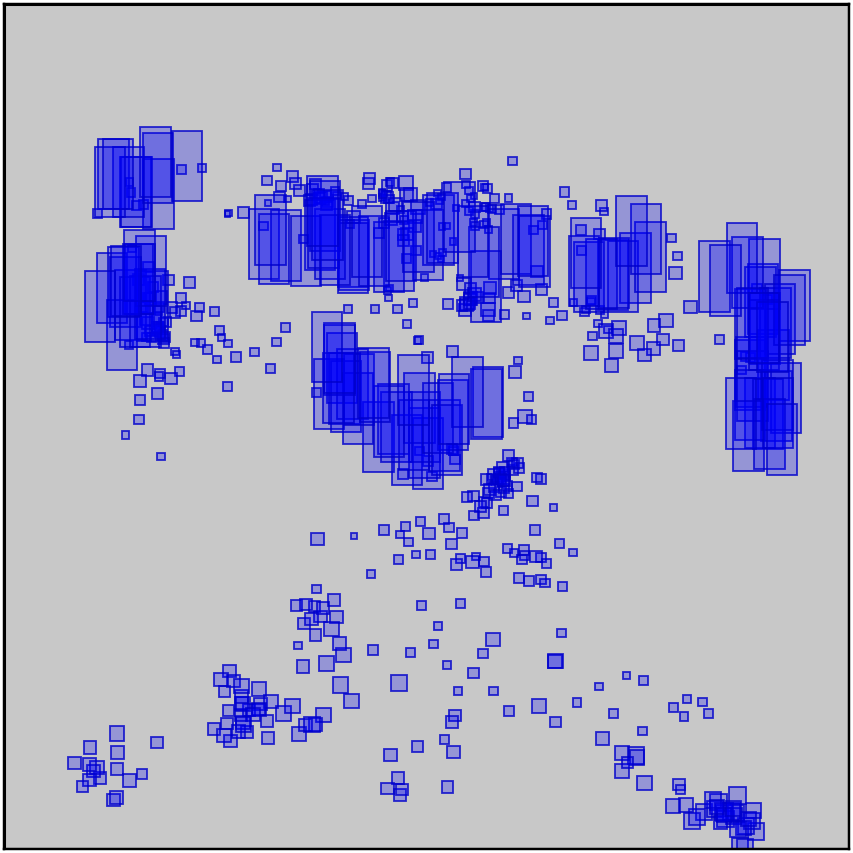} &
        \includegraphics[width=0.15\textwidth]{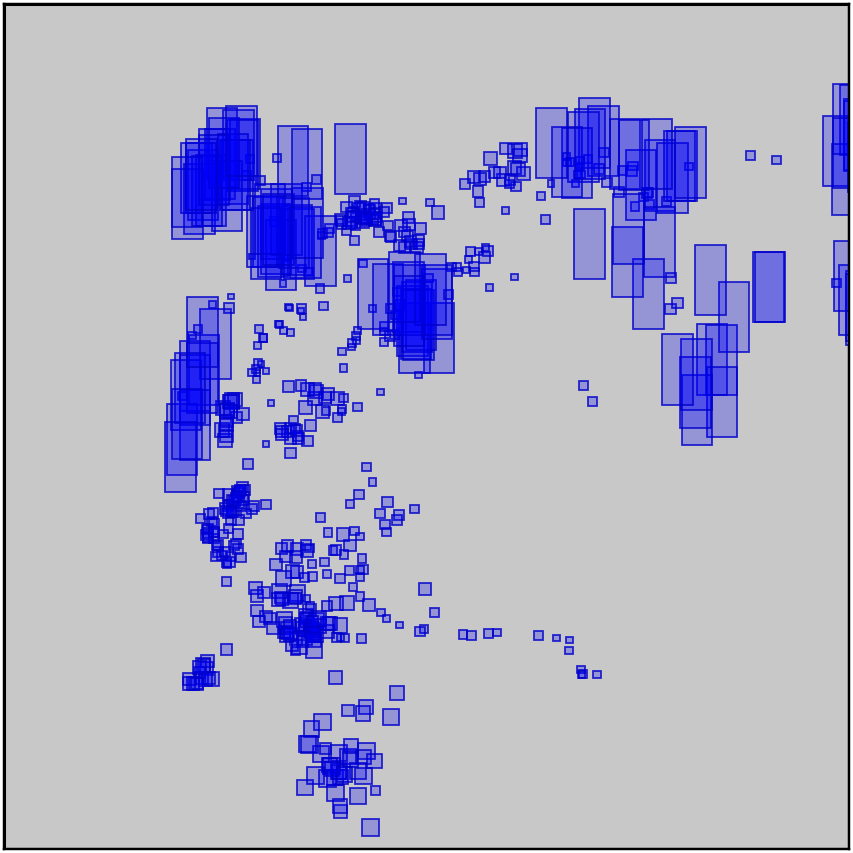} &
        \includegraphics[width=0.15\textwidth]{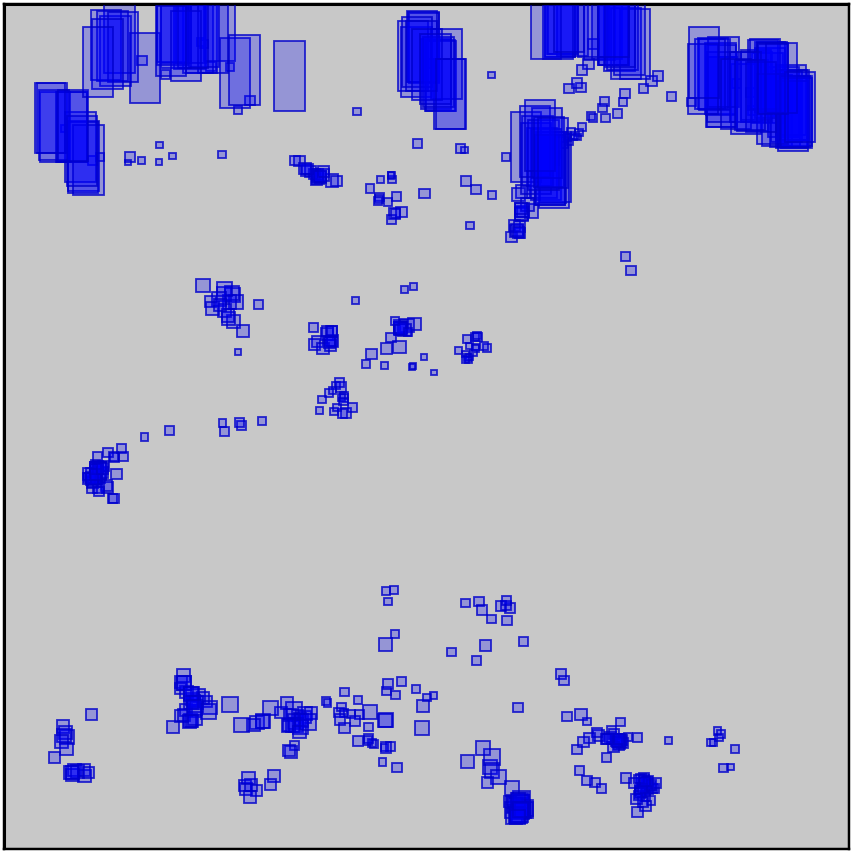} &
        \includegraphics[width=0.15\textwidth]{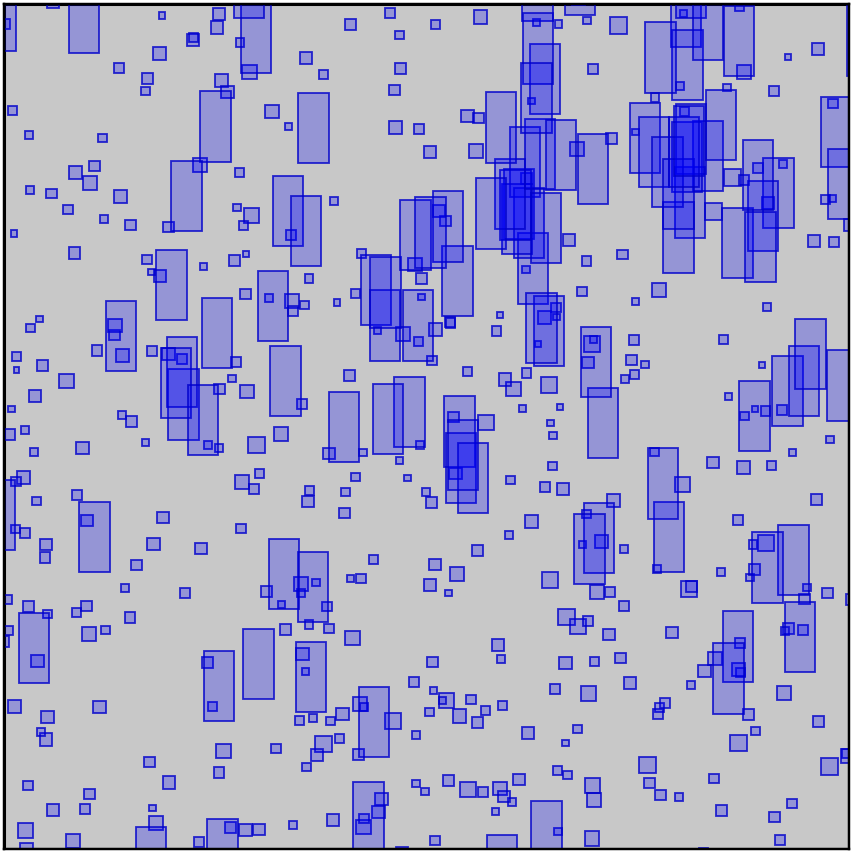} &
        \includegraphics[width=0.15\textwidth]{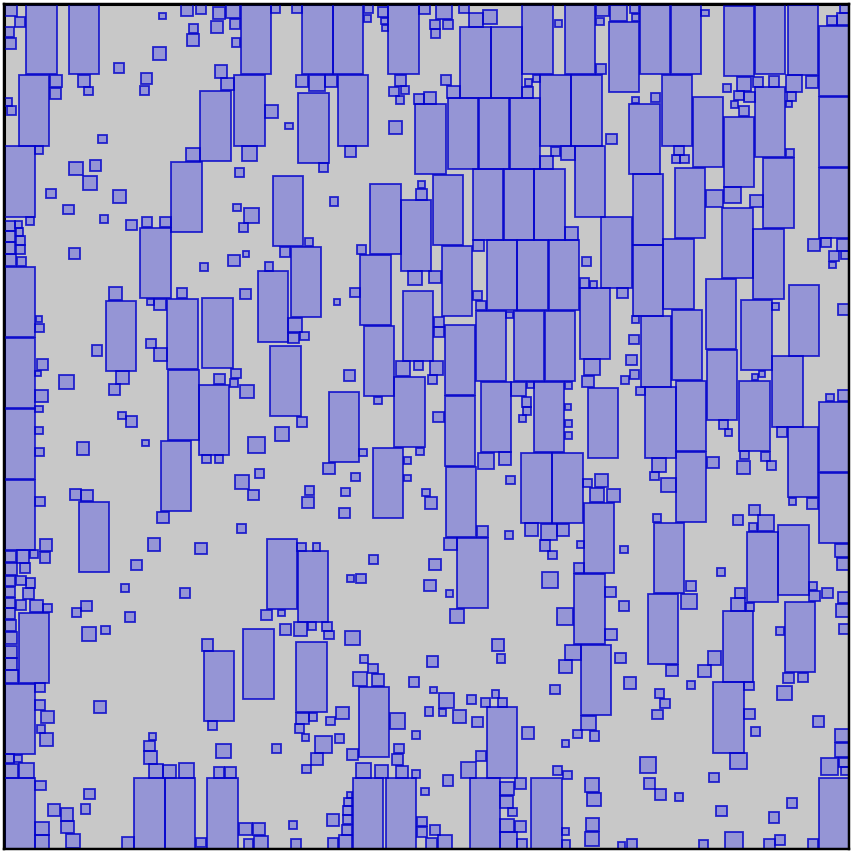}
    \end{tabular}
\caption{Visualizing the progressive denoising process on the \texttt{ariane133} benchmark. \modelname simultaneously places both SRAM macros and clustered standard-cell blocks. 
From left to right: The process initializes from Gaussian noise $x_T \sim \mathcal{N}(0, I)$ in normalized coordinates and iteratively refines the layout. Intermediate panels display the model's clean estimate $\hat{x}_0$ at selected diffusion steps, culminating in the final placement $x_0$ (rightmost panel).}
\label{fig:diffusion_process}
\end{figure*}

Figure~\ref{fig:diffusion_process} illustrates the fundamental difference between our approach and sequential RL methods. The main contributions of our work are as follows:
\begin{itemize}
    \item We propose \modelname, a generative framework that simultaneously optimizes all module positions, shifting the paradigm from sequential decision-making to parallel distribution modeling.
    \item We introduce a Vector-wise Message Passing neural backbone that captures the geometric and topological inductive biases essential for physical design.
    \item We develop a Decoupled Guidance Mechanism that effectively balances global wirelength minimization with hard physical constraints and routability.
    \item We develop a Process-Aware Data generation pipeline that synthesizes realistic netlists adhering to foundry design rules. This enables the construction of large-scale, physically compliant training datasets to overcome the scarcity of open-source industrial designs and support robust pre-training.
    \item Extensive experiments demonstrate that \modelname achieves superior generalization efficiency, delivering competitive placement quality on benchmarks while significantly reducing runtime compared to state-of-the-art learning-based baselines.
\end{itemize}

%% file: sections/relatedworked.tex
\section{Related Works}
\label{sec:relatedworks}
The evolution of chip placement methodologies has progressed through several paradigms, from classical analytical approaches to recent learning-based techniques. In this section, we examine this trajectory to position our diffusion-based method within the broader context of placement solutions.
\subsection{Classical Placement Methods}
Conventional placement approaches fall into three main categories: partitioning-based, stochastic optimization, and analytical methods.

Partitioning-based techniques \cite{DBLP:journals/tcad/RoyAPM06, DBLP:conf/ispd/KhatkhateLAYOKM04} recursively divide netlists and placement regions into manageable sub-problems. Although computationally efficient and scalable to large designs, these methods often sacrifice global optimality for runtime performance, as decisions made at higher levels constrain subsequent optimizations. The hierarchical nature of these approaches fundamentally limits their ability to find globally optimal solutions, particularly for complex and highly connected designs.

Stochastic optimization methods, particularly simulated annealing \cite{DBLP:journals/tcasI/JiangWHYY23}, dominated placement in the years ago. These approaches perform random perturbations of placements guided by a gradually decreasing temperature parameter, enabling them to escape local minima. Although SA-based placers like TimberWolf \cite{DBLP:conf/dac/SechenS86} achieved high-quality results, their prohibitive runtime complexity made them impractical for modern IC designs with millions of components.

Subsequently, analytical methods emerged as the predominant approach, with force-directed techniques \cite{DBLP:conf/aspdac/ViswanathanPC07, DBLP:journals/tcad/SpindlerSJ08} and non-linear optimizers \cite{DBLP:conf/ispd/LuZKCC16, DBLP:conf/iccad/ChenJHCC06} showing significant promise. These approaches transform discrete placement problems into continuous optimization frameworks that can be solved with gradient-based methods. Recent advances, including DREAMPlace \cite{DBLP:journals/tcad/LinJGLDRKP21}, which leverages GPU acceleration, and RePlAce \cite{DBLP:journals/tcad/ChengKKW19}, which employs an electrostatics-based formulation with Nesterov's method, have further pushed the boundaries of analytical placement. However, these methods invariably treat each design as an isolated problem, ignoring the knowledge from previous placements, a fundamental limitation that our work addresses.

\subsection{Reinforcement Learning for Placement}
The application of reinforcement learning to chip placement has gained significant traction following the breakthrough work of Mirhoseini et al. \cite{DBLP:journals/corr/abs-2004-10746}, which demonstrated that RL could surpass human experts in placement quality. Their approach, GraphPlace, represents the netlist as a graph and uses a graph neural network to learn placement policies that sequentially position macros on the chip canvas. Although revolutionary, this work exposed several limitations inherent to the RL paradigm: (1) expensive online training requirements for each new design, (2) limited generalization to unseen netlists, and (3) the compounding error problem where early placement decisions constrain later options.
Subsequent work has refined this approach while maintaining the sequential RL paradigm. DeepPR \cite{DBLP:conf/nips/ChengY21} extended GraphPlace by jointly considering placement and routing objectives, introducing a unified learning framework that produces placements with improved routability. However, it similarly requires extensive online interactions for each new design and struggles to generalize across designs.
MaskPlace \cite{DBLP:conf/nips/LaiM022} reconceptualized placement as visual representation learning, utilizing convolutional neural networks to capture spatial relationships. This approach significantly improved the handling of mixed-size macros and achieved zero-overlap placements without post-processing. The authors model the chip canvas as a 2D image and employ a policy network to place macros sequentially, guided by density and congestion masks. Despite these innovations, MaskPlace retains the fundamental limitations of the RL paradigm: expensive online training and limited generalization.
PRNet \cite{DBLP:conf/nips/ChengY21} further expanded the learning-based placement landscape by combining policy gradient methods for macro placement with generative routing networks, creating the first end-to-end neural pipeline for both placement and routing. This integration highlights the importance of considering downstream routing constraints during placement, but does not address the fundamental limitations of sequential decision making.
Despite their impressive results, all these RL-based approaches share common limitations: they require exhaustive online training for each new design, exhibit limited generalization capabilities, and follow a sequential placement paradigm that leads to compounding errors. These limitations motivate our exploration of generative models that can simultaneously place all components.

\subsection{Transfer Learning and Offline Methods}
Recent advances in transfer learning have attempted to address the generalization limitations of RL-based placers. ChiPFormer \cite{DBLP:conf/icml/LaiLTWH023} presents a significant step forward by introducing an off-line decision-transformer framework that enables knowledge transfer between designs. By pretraining on a dataset of placement examples and fine-tuning on new designs, ChiPFormer reduces training time from days to hours while maintaining competitive placement quality. 
This approach begins to bridge the gap between design-specific and generalizable placement algorithms through a novel neural architecture that effectively captures and transfers placement knowledge. However, ChiPFormer remains fundamentally bound to the sequential placement paradigm and still requires non-trivial online interactions during fine-tuning, limiting its efficiency for new designs.
In parallel, WireMask-BBO \cite{DBLP:conf/nips/Shi0L023} explored a different direction by applying black-box optimization techniques to the placement problem. Although not strictly a learning-based method, this approach demonstrates the value of global optimization strategies that consider all macro positions simultaneously rather than sequentially. However, it requires expensive search procedures for each new design and does not take advantage of cross-design knowledge.

\subsection{Alternative Learning Paradigms}
Beyond reinforcement learning, researchers have explored other learning paradigms for chip placement. Supervised learning approaches such as Flora \cite{DBLP:conf/dac/LiuJLDZWY0S22} and GraphPlanner \cite{DBLP:journals/todaes/LiuJLDZWYZS23} train neural networks to directly predict optimal placements from netlist features. These methods frame placement as a regression problem rather than a sequential decision process, allowing faster inference times. Flora employs a graph attention network to encode the netlist structure and predict macro positions, while GraphPlanner extends this approach with more sophisticated graph neural networks and loss functions. While these approaches move away from the sequential paradigm, they struggle with the fundamental challenge of generating placements that satisfy hard constraints like zero overlap, often requiring extensive post-processing. Recent work has also explored equivariant graph and hypergraph neural networks to learn structural representations of netlists, demonstrating improved generalization in EDA tasks involving connectivity and layout patterns \cite{DBLP:conf/aistats/LuoHTDRCDJ024}.

\subsection{Generative Models in EDA}
Generative models have shown promise in various electronic design automation (EDA) tasks. WellGAN \cite{DBLP:conf/dac/XuLTLSSP19} utilized generative adversarial networks for well generation in analog layouts, while ThermGAN \cite{DBLP:conf/iccad/JinSZT20} demonstrated the application of GANs to thermal map estimation. For congestion prediction, LHNN \cite{DBLP:conf/dac/WangSLHLHWLCH22} employed latent hypergraph neural networks to model complex routing patterns.
These applications highlight the potential of generative models to capture complex distributions in the EDA domain, but their application to macro placement has been limited. Unlike reinforcement learning approaches that make sequential decisions, generative models can potentially capture the joint distribution of all macro positions simultaneously, enabling more holistic placement optimization.

Our work builds on these advances by introducing a diffusion-based generative approach to chip placement. Table~\ref{tab:method_comparison} summarizes the key characteristics of recent placement approaches, highlighting differences in methodology, resolution, state space complexity, overlap guarantees, and optimization metrics. As shown, diffusion-based methods offer significant advantages over both analytical and RL-based approaches, particularly in their ability to guarantee zero overlap while maintaining high efficiency and optimizing for all key placement metrics without the exponential state-space complexity inherent in sequential methods. Unlike previous sequential methods, our approach simultaneously optimizes all macro positions through an iterative denoising process. This fundamental change in methodology eliminates the compounding errors inherent in sequential approaches while enabling effective knowledge transfer between designs through conditional generation.

\begin{table*}[t]
\caption{Comparison of placement methods across key design criteria. We evaluate representative approaches from analytical optimization, reinforcement learning (RL), offline transfer learning, and our proposed diffusion-based model (\modelname). The comparison includes placement resolution, state space complexity, overlap guarantees, efficiency in training and inference, and optimization targets such as half-perimeter wirelength (HPWL), congestion, and density. \modelname achieves simultaneous placement with zero overlap in most cases, while maintaining high efficiency and optimizing multiple placement objectives without relying on sequential decision-making or costly online training.}
\label{tab:method_comparison}
\centering
\begin{tabular}{|l|c|c|c|c|c|c|c|}
\hline
\textbf{Method} & \textbf{Family} & \textbf{Decision Paradigm} & \textbf{Search Space} & \textbf{0 \% Overlap} & \textbf{Transferability} & \textbf{Efficiency}$^1$ & \textbf{Metrics}$^2$\\
\hline
DREAMPlace \cite{DBLP:journals/tcad/LinJGLDRKP21} & Analytical & Simultaneous & Continuous & Soft Penalty & None & -/High & H,D\\
GraphPlace \cite{DBLP:journals/corr/abs-2004-10746} & RL & Sequential & Discrete Grid & No & Low & Low/Low & H,C,D \\
DeepPR \cite{DBLP:conf/nips/ChengY21} & RL & Sequential & Discrete Grid & No & Low & Med/Med & H,C \\
MaskPlace \cite{DBLP:conf/nips/LaiM022} & RL & Sequential & Discrete Grid & No & Low & Low/Low & H,C,D \\
ChiPFormer \cite{DBLP:conf/icml/LaiLTWH023} & Offline RL & Sequential & Discrete Grid & Yes$^*$ & Medium & Med/High & H,C,D \\
\modelname (Ours) & Diffusion & Simultaneous & Continuous & Yes & High & High/High & H,C,D \\
\hline
\multicolumn{8}{l}{$^1$ Training/Inference efficiency} \\
\multicolumn{8}{l}{$^2$ H = Half-Perimeter Wire Length, C = Congestion, D = Density} \\
\multicolumn{8}{l}{$^3$ ChiPFormer and Ours achieves zero overlap on most benchmarks but not guaranteed on all circuits (3.27\% overlap reported)} \\
\end{tabular}
\end{table*}

\begin{figure}[t]
    \centering
    \subfloat[Online RL Placement]{
        \includegraphics[width=0.95\columnwidth]{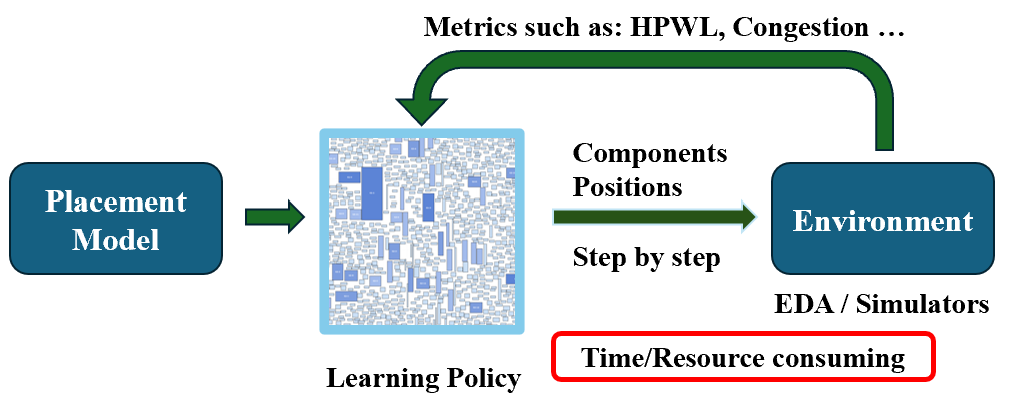}
        \label{fig:rl_pipeline}
    }
    \hfill 
    \subfloat[Diffusion (Ours)]{
        \includegraphics[width=0.95\columnwidth]{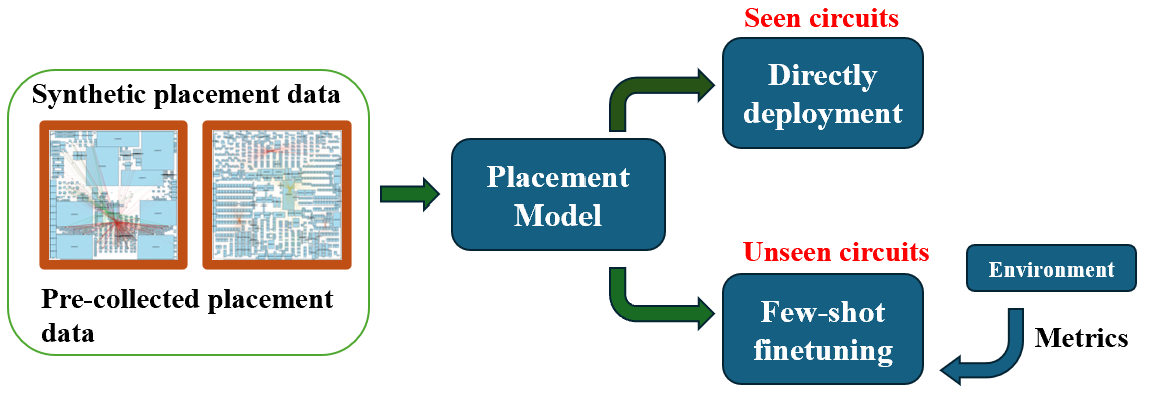}
        \label{fig:diffplace_pipeline}
    }
    
    \caption{\textbf{Paradigm comparison between (a) Online RL and (b) \modelname.} 
    RL approaches (e.g., GraphPlace, MaskPlace) suffer from high computational costs due to iterative interactions with EDA tools during online training for each new circuit. 
    In contrast, our diffusion framework learns a transferable policy offline, allowing for efficient one-pass generation (inference) on unseen netlists without retraining.}
    \label{fig:pipeline_comparison}
\end{figure}

To visually underscore the efficiency gap discussed above, Fig. \ref{fig:pipeline_comparison} contrasts the operational paradigms of RL-based placers and our diffusion framework. While RL methods rely on computationally expensive, per-design environment interactions to learn placement policies from scratch, \modelname leverages offline pre-training on diverse datasets. This paradigm shift eliminates the need for online retraining, enabling rapid zero-shot inference on unseen netlists

\subsection*{Comparison with Concurrent Work}
A concurrent study by Lee et al. \cite{DBLP:conf/icml/LeeNEDAW25} also explores diffusion for placement but diverges fundamentally from our approach in two key aspects. First, regarding architecture, while \cite{DBLP:conf/icml/LeeNEDAW25} relies on scalar GNNs with sinusoidal positional encodings, \modelname{} employs vector-wise message passing. This design captures geometric vectors ($\mathbf{x}_j - \mathbf{x}_i$), ensuring translation invariance essential for physical design. Second, unlike the universal guidance in \cite{DBLP:conf/icml/LeeNEDAW25} which computes gradients for all objectives, we introduce a decoupled guidance mechanism. We optimize global wirelength implicitly via energy conditioning ($E_{rel}$) and reserve explicit gradient injection solely for local physical constraints (overlap/congestion), avoiding the computational instability associated with deriving gradients for global metrics on noisy manifolds.

%% file: sections/preliminaries.tex
\section{Preliminaries}
\label{sec:preliminaries}

\subsection{Chip Placement Problem Formulation}
We formulate the chip placement problem as finding the optimal spatial configuration for a set of circuit modules. Let $G = (V, E)$ denote the netlist hypergraph, where $V = \{v_1, \dots, v_n\}$ is the set of modules and $E = \{e_1, \dots, e_m\}$ is the set of nets. Each module $v_i$ has width $w_i$ and height $h_i$.

We define the placement state as a vector $\mathbf{x} \in \mathbb{R}^{2n}$, where $\mathbf{x}_{2i-1}$ and $\mathbf{x}_{2i}$ correspond to the geometric center coordinates $(x_i, y_i)$ of module $v_i$ on a continuous canvas region $\mathcal{R} = [0, W] \times [0, H]$. The goal is to determine a placement $\mathbf{x}^*$ that minimizes a composite objective function $\mathcal{J}(\mathbf{x}, G)$ subject to physical constraints:
\begin{equation}
    \mathbf{x}^* = \arg\min_{\mathbf{x}} \mathcal{J}(\mathbf{x}, G) \quad \text{s.t.} \quad
    \begin{cases}
        \mathbf{x}_i \in [0, W] \times [0, H], & \forall v_i \in V \quad \\
        \text{Overlap}(\mathbf{x}) \to 0, & \quad  \\
        \text{Density}_g(\mathbf{x}) \le \rho_{\text{target}}, & \forall g \in \mathcal{G} \quad 
    \end{cases}
    \label{eq:problem_formulation}
\end{equation}
where $\mathcal{G}$ represents the set of discretized grid bins covering the canvas area.

\subsection{Optimization Objectives}
The objective function $\mathcal{J}$ typically comprises three key physical design metrics, which we define formally below. These definitions serve as the basis for the energy functions used in our diffusion guidance.



\subsubsection{Half-Perimeter Wirelength (HPWL)}
HPWL is the primary proxy for wirelength and timing. For a net $e \in E$, let $P_e$ be the set of pins connected by $e$. The absolute position of a pin $p \in P_e$ belonging to module $v_{i(p)}$ is given by $(x_{i(p)} + \delta x_p, y_{i(p)} + \delta y_p)$, where $(\delta x_p, \delta y_p)$ denotes the relative pin offset from the module center. The HPWL of net $e$ is defined as:
\begin{equation}
\begin{split}
    \text{HPWL}_e(\mathbf{x}) &= \left( \max_{p \in P_e} (x_{i(p)} + \delta x_p) - \min_{p \in P_e} (x_{i(p)} + \delta x_p) \right) \\
    &\quad + \left( \max_{p \in P_e} (y_{i(p)} + \delta y_p) - \min_{p \in P_e} (y_{i(p)} + \delta y_p) \right).
\end{split}
\end{equation}
The total wirelength is $\text{HPWL}(\mathbf{x}) = \sum_{e \in E} \text{HPWL}_e(\mathbf{x})$. Since the $\max/\min$ operators are non-differentiable, we employ the weighted-average (WA) smoothing technique during gradient-based guidance steps.

\subsubsection{Routing Congestion}
To estimate routability, we utilize the RUDY (Rectangular Uniform Wire DensitY) estimator. The chip canvas is discretized into a grid $\mathcal{G}$. For a grid tile $g \in \mathcal{G}$, the congestion is modeled as:
\begin{equation}
    \text{Congestion}(g; \mathbf{x}) = \sum_{e \in E} \mathbb{I}(g \cap \text{BBox}_e(\mathbf{x})) \cdot \frac{1}{w_e + h_e},
\end{equation}
where $\text{BBox}_e$ is the bounding box of net $e$, and $\mathbb{I}$ is the indicator function.

\subsubsection{Overlap Constraint}
Strict legality requires disjoint module areas. We quantify the violation using the total pairwise overlap area:
\begin{equation}
    \text{Overlap}(\mathbf{x}) = \sum_{i \neq j} \max(0, \Delta x_{ij}) \cdot \max(0, \Delta y_{ij}),
\end{equation}
where $\Delta x_{ij} = \min(x_i + \frac{w_i}{2}, x_j + \frac{w_j}{2}) - \max(x_i - \frac{w_i}{2}, x_j - \frac{w_j}{2})$ represents the overlap width (similarly for height $\Delta y_{ij}$). In our framework, we treat this as a hard constraint to be minimized via the generative process, aiming for $\text{Overlap}(\mathbf{x}) \to 0$.

\subsection{Denoising Diffusion Probabilistic Models (DDPM)}
DDPMs are a class of generative models that learn to reverse a Markovian diffusion process.
\subsubsection{Forward Process}
Given a data sample $\mathbf{x}_0$ from distribution $q(\mathbf{x})$, the forward process produces a sequence of latent variables $\mathbf{x}_1, \dots, \mathbf{x}_T$ by adding Gaussian noise:
\begin{equation}
    q(\mathbf{x}_t | \mathbf{x}_{t-1}) = \mathcal{N}(\mathbf{x}_t; \sqrt{1-\beta_t}\mathbf{x}_{t-1}, \beta_t\mathbf{I}),
\end{equation}
where $\beta_t$ is the noise schedule. A closed-form property allows sampling $\mathbf{x}_t$ directly from $\mathbf{x}_0$:
\begin{equation}
    q(\mathbf{x}_t | \mathbf{x}_0) = \mathcal{N}(\mathbf{x}_t; \sqrt{\bar{\alpha}_t}\mathbf{x}_0, (1-\bar{\alpha}_t)\mathbf{I}), \quad \text{with } \bar{\alpha}_t = \prod_{s=1}^t (1-\beta_s).
\end{equation}

\subsubsection{Reverse Process}
The generative process learns to invert the noise by estimating the posterior $q(\mathbf{x}_{t-1}|\mathbf{x}_t)$. This is parameterized as:
\begin{equation}
    p_\theta(\mathbf{x}_{t-1} | \mathbf{x}_t) = \mathcal{N}(\mathbf{x}_{t-1}; \boldsymbol{\mu}_\theta(\mathbf{x}_t, t), \boldsymbol{\Sigma}_\theta(\mathbf{x}_t, t)).
\end{equation}
The model is trained to predict the noise $\boldsymbol{\epsilon}$ via a network $\boldsymbol{\epsilon}_\theta(\mathbf{x}_t, t)$, optimizing the simplified objective:
\begin{equation}
    \mathcal{L}(\theta) = \mathbb{E}_{t \sim \mathcal{U}(0,T), \mathbf{x}_0 \sim p_{\text{data}}, \boldsymbol{\epsilon} \sim \mathcal{N}(\mathbf{0, I})} \left[ \|\boldsymbol{\epsilon} - \boldsymbol{\epsilon}_\theta(\mathbf{x}_t, t)\|^2 \right].
\end{equation}
In Section \ref{sec:methodology}, we adapt this generic framework to the conditional generation of chip placements subject to the constraints defined in Eq. (\ref{eq:problem_formulation}).

%% file: sections/methodology.tex
\section{Methodology} 
\label{sec:methodology}

\subsection{Overview of \modelname}
\begin{figure*}[h]
    \centering
    \includegraphics[width=0.95\linewidth]{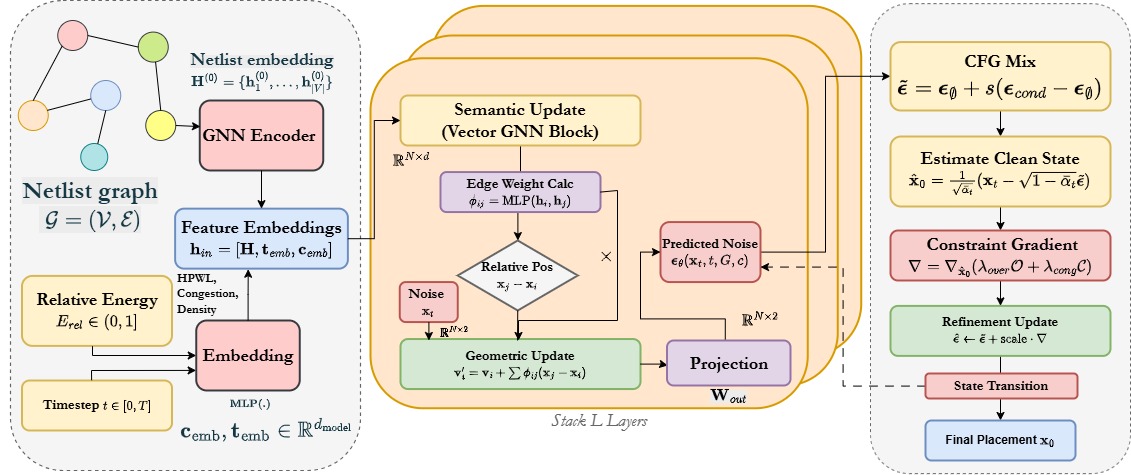}
   \caption{\textbf{Overview of the DiffPlace framework.} 
(A) The model conditions on the netlist graph $\mathcal{G}$ and relative energy targets. 
(B) The score network employs Vector-wise Message Passing to capture relative geometric dependencies. 
(C) The reverse process integrates Manifold Constraint Guidance: $\hat{\mathbf{x}}_0$ is estimated to compute physical constraint gradients, iteratively resolving overlaps during generation.}
\label{fig:framework}
    \label{fig:framework_overview}
\end{figure*}

We propose \modelname, a conditional diffusion framework that generates placement coordinates $\mathbf{x}$ by reversing the diffusion process conditioned on the netlist graph $G$ and explicit quality constraints. As illustrated in Fig.~\ref{fig:framework_overview}, our approach enables simultaneous optimization of all modules, contrasting with sequential RL methods.

\subsection{Conditional Diffusion for Netlists}
We extend the standard DDPM to a conditional distribution $p_\theta(\mathbf{x} | G, c)$. The reverse process is defined as:
\begin{equation}
    p_\theta(\mathbf{x}_{t-1} | \mathbf{x}_t, G, c) = \mathcal{N}(\mathbf{x}_{t-1}; \boldsymbol{\mu}_\theta(\mathbf{x}_t, t, G, c), \sigma_t^2 \mathbf{I}).
\end{equation}
where $\sigma_t^2$ is set to $\beta_t$ following the standard DDPM schedule \cite{DBLP:conf/nips/HoJA20}, stabilizing the training process.
The mean $\boldsymbol{\mu}_\theta$ is derived from the predicted noise $\boldsymbol{\epsilon}_\theta$:
\begin{equation}
    \boldsymbol{\mu}_\theta(\mathbf{x}_t, t, G, c) = \frac{1}{\sqrt{\alpha_t}}\left( \mathbf{x}_t - \frac{\beta_t}{\sqrt{1-\bar{\alpha}_t}} \boldsymbol{\epsilon}_\theta(\mathbf{x}_t, t, G, c) \right).
\end{equation}

Here, $c$ represents the relative energy condition, which steers the generation toward high-quality regions of the solution space defined in Section II.

\subsection{Energy-Conditioned Mechanism}
A critical innovation of \modelname{} is the energy-based conditioning that allows zero-shot generalization to unseen netlists.

\subsubsection{Composite Energy Function}
Leveraging the metrics defined in Preliminaries, we construct a differentiable energy function $E(\mathbf{x}, G)$:
\begin{equation}
\begin{aligned}
E(\mathbf{x}, G) &=
\lambda_{\text{HPWL}} \frac{\text{HPWL}(\mathbf{x})}{L_{\text{norm}}}
+ \lambda_{\text{cong}} \text{RUDY}(\mathbf{x}, G) \\
&\quad + \lambda_{\text{over}} \text{Overlap}(\mathbf{x})
\end{aligned}
\label{eq:energy_function}
\end{equation}

where $\lambda$ terms are weighting factors determined empirically on validation sets. $L_{\text{norm}}$ serves as a normalization constant to ensure the HPWL term is scale-invariant. We define $L_{\text{norm}}$ as the theoretical upper bound of the total wirelength estimate:
$$L_{\text{norm}} = |E_{\text{net}}| \cdot (W_{\text{die}} + H_{\text{die}})$$

$|E_{\text{net}}|$ is the total number of nets, and $W_{\text{die}}, H_{\text{die}}$ are the width and height of the placement region, respectively.

\subsubsection{Relative Quality Normalization}
Absolute energy values vary significantly across circuit designs due to different instance counts. To ensure the condition $c$ is scale-invariant, we introduce the \textit{Relative Energy Score} ($E_{\text{rel}}$):
\begin{equation}
    E_{\text{rel}}(\mathbf{x}, G) = \exp\left( - \kappa \cdot \frac{E(\mathbf{x}, G) - E_{\text{ana}}(G)}{E_{\text{ana}}(G)} \right).
    \label{eq:relative_energy}
\end{equation}
Here, $E_{\text{ana}}(G)$ represents the analytical lower bound of the wirelength, computed via spectral placement or quadratic relaxation. This normalization provides a stable reference point, ensuring $E_{\text{rel}} \in (0, 1]$ regardless of circuit size, where $\kappa > 0$ is a decay rate parameter controlling the sensitivity of the condition scale

\subsubsection{Training Objective}
The network is trained to predict the added noise $\boldsymbol{\epsilon}$, conditioned on the netlist $G$ and its corresponding quality score:
\begin{equation}
    \mathcal{L}(\theta) = \mathbb{E}_{t, \mathbf{x}_0, \boldsymbol{\epsilon}, G} \left[ \| \boldsymbol{\epsilon} - \boldsymbol{\epsilon}_\theta(\mathbf{x}_t, t, G, E_{\text{rel}}(\mathbf{x}_0, G)) \|^2 \right].
\end{equation}
During inference, we set the target condition $E_{\text{rel}} \to 1$ to guide the diffusion trajectory toward the optimal manifold.

\subsection{Neural Architecture}
The score network $\boldsymbol{\epsilon}_\theta$ employs a Vector-wise Graph Neural Network (GNN) encoder to capture netlist topology.
To explicitly model geometric relationships critical for placement, we utilize a Vector-wise Message Passing scheme. Let $\mathbf{h}_i^{(l)}$ be the feature of node $v_i$. The update rule aggregates relative position vectors:
\begin{equation}
    \mathbf{h}_i^{(l+1)} = \text{Update}\left( \mathbf{h}_i^{(l)}, \sum_{j \in \mathcal{N}(i)} \phi(\mathbf{e}_{ij}) (\mathbf{x}_j - \mathbf{x}_i) \right),
\end{equation}
where $\phi(\mathbf{e}_{ij}) \in \mathbb{R}$ is a learnable scalar edge weight derived from net features. 
By explicitly encoding relative geometric vectors $(\mathbf{x}_j - \mathbf{x}_i)$, this formulation enforces translation invariance—a fundamental requirement for global placement optimization. Unlike absolute position encodings, this approach allows the model to learn robust spatial dependencies directly from the noisy state $\mathbf{x}_t$, ensuring generalization across different die regions.

\subsection{Constraint-Aware Guided Sampling}
To strictly enforce the overlap constraint (Eq. 6) and minimize congestion (Eq. 5), we employ a hybrid guidance strategy during inference.

\subsubsection{Classifier-Free Guidance (CFG)}
We effectively amplify the high-quality condition using CFG:
\begin{equation}
    \tilde{\boldsymbol{\epsilon}}_\theta = \boldsymbol{\epsilon}_\theta(\mathbf{x}_t, t, \emptyset) + s_{\text{cfg}} \cdot (\boldsymbol{\epsilon}_\theta(\mathbf{x}_t, t, G, E_{\text{rel}}=1) - \boldsymbol{\epsilon}_\theta(\mathbf{x}_t, t, \emptyset)),
\end{equation}
where $s_{\text{cfg}}$ is the guidance scale.

\subsubsection{Manifold Constraint Guidance}
Since $E_{\text{rel}}$ is a scalar summary, it acts as a global guide but may lack the granularity to resolve local overlaps. We introduce an explicit gradient guidance term to address this. To avoid the computational prohibitive cost of backpropagating through the U-Net architecture \cite{DBLP:conf/miccai/RonnebergerFB15}, we first estimate the clean state $\hat{\mathbf{x}}_0$ from the current noisy state $\mathbf{x}_t$ using the DDPM reparameterization:
\begin{equation}
    \hat{\mathbf{x}}_0(\mathbf{x}_t) = \frac{1}{\sqrt{\bar{\alpha}_t}} (\mathbf{x}_t - \sqrt{1-\bar{\alpha}_t}\tilde{\boldsymbol{\epsilon}}_\theta(\mathbf{x}_t)).
\end{equation}
We then define a constraint loss $\mathcal{L}_{\text{cons}} = \lambda_{\text{over}}\text{Overlap}(\hat{\mathbf{x}}_0) + \lambda_{\text{cong}}\text{Congestion}(\hat{\mathbf{x}}_0)$. The guidance gradient with respect to $\mathbf{x}_t$ is derived via the chain rule:
\begin{equation}
    \nabla_{\mathbf{x}_t} \mathcal{L}_{\text{cons}} = \nabla_{\hat{\mathbf{x}}_0} \mathcal{L}_{\text{cons}} \cdot \frac{\partial \hat{\mathbf{x}}_0}{\partial \mathbf{x}_t}.
\end{equation}

Computing the exact Jacobian $\frac{\partial \hat{\mathbf{x}}_0}{\partial \mathbf{x}_t}$ requires backpropagating through the entire U-Net denoiser, which is computationally prohibitive for high-dimensional placement states. Although the denoiser is non-linear, the local linearity assumption holds sufficiently for small timesteps, allowing efficient gradient computation without full backpropagation. Following established approximations in constrained diffusion \cite{DBLP:conf/iclr/ChungKMKY23, DBLP:conf/nips/DhariwalN21}, we assume the denoiser is locally linear and approximate the Jacobian as a scalar scaling factor: $\frac{\partial \hat{\mathbf{x}}_0}{\partial \mathbf{x}_t} \approx \frac{1}{\sqrt{\bar{\alpha}_t}}\mathbf{I}$.
Consequently, the modified score estimate incorporates the physical gradient as:
\begin{equation}
    \hat{\boldsymbol{\epsilon}}_\theta(\mathbf{x}_t, t) = \tilde{\boldsymbol{\epsilon}}_\theta(\mathbf{x}_t, t) + s_{\text{cons}} \sqrt{1-\bar{\alpha}_t} \nabla_{\hat{\mathbf{x}}_0} \mathcal{L}_{\text{cons}}(\hat{\mathbf{x}}_0),
\end{equation}
where $s_{\text{cons}}$ is a time-dependent guidance scale. This approximation effectively projects the gradient of the physical constraints (computed on the estimated clean manifold $\hat{\mathbf{x}}_0$) back onto the noise space $\mathbf{x}_t$, acting as a restorative force—shifting the denoised output opposite to the constraint gradient—that iteratively pushes modules out of invalid configurations (e.g., overlaps) during the sampling trajectory. 

The inference complexity of \modelname is $O(T \cdot (N_{nodes} + N_{edges}))$, where $T$ is the number of diffusion steps. This is linear with respect to the circuit size, enabling scalability to large netlists. Grid-based density estimation is implemented for the overlap and congestion constraints. Macro areas are scattered onto a fixed $M \times M$ grid using efficient scatter-add operations. This reduces the computational complexity of the constraint gradient $\nabla \mathcal{L}_{\text{cons}}$ from quadratic $\mathcal{O}(N_{\text{macro}}^2)$ to linear $\mathcal{O}(N_{\text{nodes}} + M^2)$. In contrast, sequential RL methods \cite{DBLP:conf/nips/LaiM022, DBLP:journals/corr/abs-2004-10746} typically scale as $O(N_{macros} \cdot \text{GridSize})$, where the sequential dependency prevents parallelization. Analytical methods \cite{DBLP:conf/ispd/LuZKCC16} scale with the number of solver iterations, which can degrade significantly for complex constraints. Our fixed-step generative process ensures predictable and efficient runtime.

\begin{algorithm}[h]
\caption{Diffplace Inference}
\label{alg:inference_hdlp}
\begin{algorithmic}[1]
\Require Netlist $G$, Pre-trained Diffusion Model $\boldsymbol{\epsilon}_\theta$, Die Specs $D$, Guidance Scale $s(t)$
\Ensure Optimized \& Legalized Placement $\mathbf{P}_{final}$

\State \textbf{Global Placement}
\State Sample $\mathbf{x}_T \sim \mathcal{N}(\mathbf{0}, \mathbf{I})$
\For{$t = T, \dots, 1$}
    \State $\hat{\boldsymbol{\epsilon}} \leftarrow \boldsymbol{\epsilon}_\theta(\mathbf{x}_t, t, G)$
    \State Estimate denoised observation: $\mathbf{x}_{prev} \leftarrow \text{DDIM\_Step}(\mathbf{x}_t, \hat{\boldsymbol{\epsilon}}, t)$

    \If{$t < T_{guide}$}
        \State Compute Density Map: $\boldsymbol{\rho} = \text{Conv}_{Gauss}(\sum_{v \in V} \mathbb{I}(v \text{ at } \mathbf{x}_{prev}))$
        \State Compute Overflow Loss: $\mathcal{L}_{den} = ||\text{ReLU}(\boldsymbol{\rho} - \rho_{target})||^2_2$
        \State Inject Gradient: $\mathbf{x}_{prev} \leftarrow \mathbf{x}_{prev} - s(t) \cdot \nabla_{\mathbf{x}_{prev}} \mathcal{L}_{den}$
    \EndIf
    \State $\mathbf{x}_{t-1} \leftarrow \mathbf{x}_{prev}$
\EndFor
\State $\mathbf{P}_{raw} \leftarrow \text{Denormalize}(\mathbf{x}_0)$

\State \textbf{SequencePairLegalizer}
\State $(S_+, S_-) \leftarrow \text{ExtractSequencePair}(\mathbf{P}_{raw})$
\State Construct Horizontal Constraint Graph $G_h(V, E_h)$ based on $(S_+, S_-)$
\State Construct Vertical Constraint Graph $G_v(V, E_v)$ based on $(S_+, S_-)$
\State Solve LP/Longest Path:
\State \quad $x_j \ge x_i + w_i, \quad \forall (i, j) \in E_h$
\State \quad $y_j \ge y_i + h_i, \quad \forall (i, j) \in E_v$
\State $\mathbf{P}_{legal} \leftarrow \text{UpdatePositions}(\mathbf{P}_{raw}, G_h, G_v)$

\State \textbf{TetrisLegalizer}
\State $\mathbf{P}_{final} \leftarrow \text{GreedyGridPacking}(\mathbf{P}_{legal}, \text{GridRes})$
\State \textbf{return} $\mathbf{P}_{final}$
\end{algorithmic}
\end{algorithm}

\subsection{Process-Aware Data Generation}
To facilitate pre-training, we generate synthetic netlists that adhere to realistic foundry rules (e.g., 45nm process constraints $\mathbf{P}_{45}$). The generation process, detailed in Algorithm \ref{alg:hierarchical_generation}, ensures that the synthetic graphs exhibit degree distributions and clustering coefficients similar to real ISPD benchmarks.



\begin{algorithm}[t]
\caption{Physics-Aware Hierarchical Netlist Generation}
\label{alg:hierarchical_generation}
\begin{algorithmic}[1]
\Require Target specs $(N_{total}, A_{target})$, Process params $\mathbf{P}_{45}$
\Ensure Netlist $G = (V, E)$, Placement $\mathbf{x}$

\State \textbf{Module Generation \& Initial Placement}
\State $N_{macro} \sim \text{Poisson}(\lambda_{macro} \cdot \log N_{total})$
\State Generate macro areas $A_{macro}$ via LogNormal distribution
\State Initialize $\mathbf{x}_{macro}$ via spectral embedding constrained to canvas boundary
\State Partition remaining area $\mathcal{R}$ via Voronoi tessellation for std cells

\State \textbf{Physics-Constrained Edges}
\For{each potential connection $(i,j)$}
    \State Compute Manhattan distance $d_{ij} = \|\mathbf{x}_i - \mathbf{x}_j\|_1$
    \State Calculate probability $p_{ij} = p(\text{edge}_{ij} \mid d_{ij}, \mathbf{P}_{45})$
    \State Add edge $e_{ij}$ to $E$ with probability $p_{ij}$
\EndFor

\State \textbf{Validation}
\State $valid \leftarrow \text{False}$
\For{$attempt = 1$ \textbf{to} $MAX\_RETRIES$}
    \State Verify Rent's rule compliance and routing capacity
    \If{metrics meet criteria}
        \State $valid \leftarrow \text{True}$
        \State \textbf{break}
    \EndIf
    \State Adjust generation parameters (e.g., relax clustering)
\EndFor

\If{$valid$ is \text{False}}
    \State \textbf{Return} Failure
\Else
    \State \Return $(G, \mathbf{x})$
\EndIf
\end{algorithmic}
\end{algorithm}

To facilitate pre-training, we generate synthetic netlists that adhere to realistic foundry rules. Specifically, to ensure physical routability matching industrial standards, we enforce a target utilization density $\rho_{target} \in [0.7, 0.85]$ during the module generation phase. The generation process, detailed in Algorithm \ref{alg:hierarchical_generation}. To empirically validate the fidelity of our generative process, we analyze the topological characteristics of the generated synthetic netlists. We provide additional statistical visualizations of the generated data in the Appendix. 

\footnote{\textbf{Our benchmark-based data samples and checkpoint are shared on }
\href{https://drive.google.com/drive/folders/1SlZPJYS9f-WEMCIA__49vGuGEVeFxAnI?usp=drive_link}{Google Drive}.}

%% file: sections/experiments.tex
\section{Experiments}
\label{sec:experiments}

\subsubsection{Experimental Setup}
Our models are implemented using Pytorch and PyTorch Geometric Deep Learning Frameworks, trained on machines with Intel Xeon Gold CPUs, using a single Nvidia RTX5880 Ada GPU. We train our models using Adam Optimizer \cite{DBLP:journals/corr/KingmaB14} with the following hyperparameters:
\begin{itemize}
    \item {Learning rate}: $3 \times 10^{-4}$ with cosine annealing schedule, decaying to $1 \times 10^{-6}$
    \item {Adam parameters}: $\beta_1 = 0.9$, $\beta_2 = 0.999$, $\epsilon = 1 \times 10^{-8}$
    \item {Training steps}: 50K steps for pre-training
    \item {Batch size}: 32 for training, 16 for inference
    \item {Gradient clipping}: Clip norm at 1.0
    \item {Diffusion steps}: $T = 50$ during inference with DDIM sampling
    \item {EMA decay}: 0.9999 for model weights
\end{itemize}

\subsubsection{Benchmarks and Datasets}
To ensure a comprehensive evaluation across different design eras and complexities, we evaluated \modelname on two distinct categories of benchmarks:
\textbf{Academic Benchmarks:} We utilized ISPD05 \cite{DBLP:conf/ispd/NamAVWY05} and the ICCAD04 \cite{DBLP:conf/iccad/AdyaCRPM04}. These benchmarks have served as the standard baselines for placement research for decades, offering a wide range of circuit complexities. \textbf{Industrial Benchmarks:} we incorporated real-world open-source circuits from the TILOS benchmark \cite{DBLP:conf/ispd/ChengKKWW23}. This set includes Ariane (RISC-V CPU), Mempool-tile, and NVDLA. These designs feature complex datapath structures and high macro utilization, representing the challenges of modern physical design. We provide statistics of the Academic and Industrial Benchmark Circuits in the Appendix. 

To ensure generalization and prevent data leakage, we enforce a Zero-Shot Evaluation Protocol by partitioning our data into disjoint sets ($D_{train} \cap D_{test} = \emptyset$). The model is pre-trained solely on a composite dataset of:

\begin{itemize}
    \item \textbf{Process-Aware Synthetic Netlists}: A corpus of 1,000 procedural designs adhering to Rent’s Rule and power-law degree distributions. This dataset incorporates diverse variations in macro count ($50\text{--}2000$) and connectivity (fanout $2\text{--}50$) to enable the model to learn fundamental topological priors without bias.
    
    \item \textbf{Augmented Reference Data}: We utilize the augmented offline dataset provided by ChiPFormer~\cite{DBLP:conf/icml/LaiLTWH023}. Each file contains 100 placement results from one circuit generated by a trained MaskPlace model. When evaluating on ISPD05, all circuits belonging to that benchmark are removed from the training corpus.
\end{itemize}

Consequently, academic benchmarks and modern RISC-V designs reported in our experiments are reserved exclusively for inference. This ensures the model optimizes \textit{unseen} topologies based solely on generalized physical laws.

\subsubsection{Evaluation Metrics}
Following standard practices in chip placement evaluation \cite{DBLP:journals/tcad/LinJGLDRKP21, DBLP:conf/nips/LaiM022}, we utilize the following metrics:
\begin{itemize}
    \item \textbf{HPWL}: Half-perimeter wirelength, which serves as a proxy for actual wirelength.
    \item \textbf{Congestion}: Maximum routing congestion across the chip canvas, measured using the RUDY estimator.
    \item \textbf{Overlap Ratio}: Percentage of overlap area between modules.
    \item \textbf{Runtime}: Wall-clock time required for placement, measured in minutes.
\end{itemize}

\subsection{Macro Placement Results}
\subsubsection{Performance on academic benchmark}

Table ~\ref{tab:ispd_macro_hpwl} and \ref{tab:ibm_macro_hpwl} summarize the wirelength performance across academic benchmarks. On smaller benchmarks (\texttt{ibm01}--\texttt{ibm04}, \texttt{adaptec1-2}--\texttt{bigblue1}), search-based approaches (WireMask-BBO) and RL baselines exhibit a marginal advantage. This performance gap is intrinsic to the algorithmic paradigm: in restricted combinatorial spaces, iterative search methods can exhaustively explore discrete moves to locate exact solutions. In contrast, DiffPlace operates as a continuous generative model, where the stochastic nature of the reverse diffusion process yields a high-fidelity approximation of the data distribution but lack the position fine-tuning capability of discrete solvers in low-dimensional manifolds. The results reveal the advantage of \modelname in high-complexity regimes, particularly on the large-scale \texttt{bigblue} and \texttt{ibm} series. This indicates that the learned gradient fields successfully guide the sampling process towards global energy minima, avoiding the local optima traps inherent in heuristic search. 

\begin{figure}[t]
    \centering
    \subfloat[GraphPlace]{\includegraphics[width=0.49\linewidth]{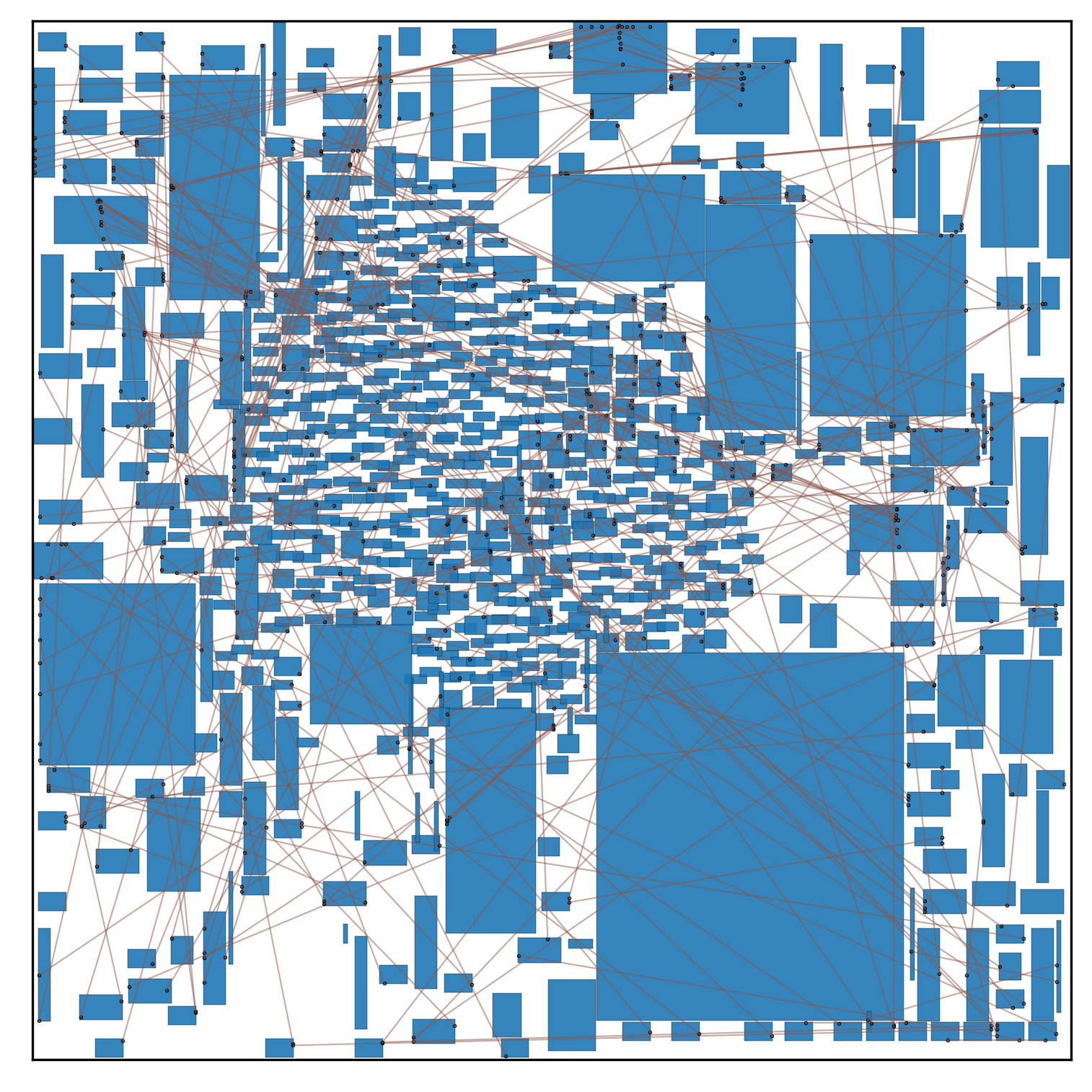}} \hfill
    \subfloat[DREAMPlace]{\includegraphics[width=0.49\linewidth]{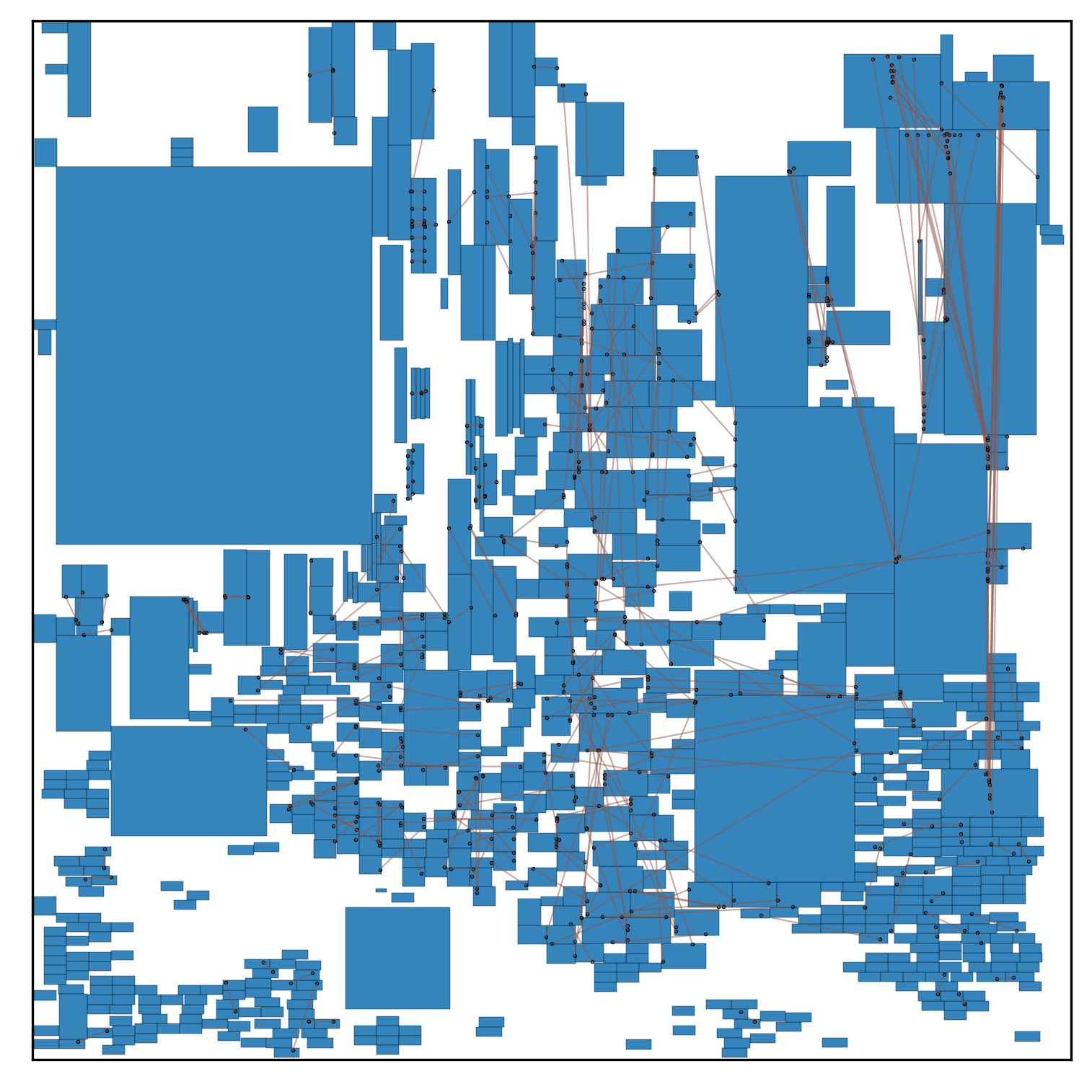}} \\
    \subfloat[DeepPR]{\includegraphics[width=0.49\linewidth]{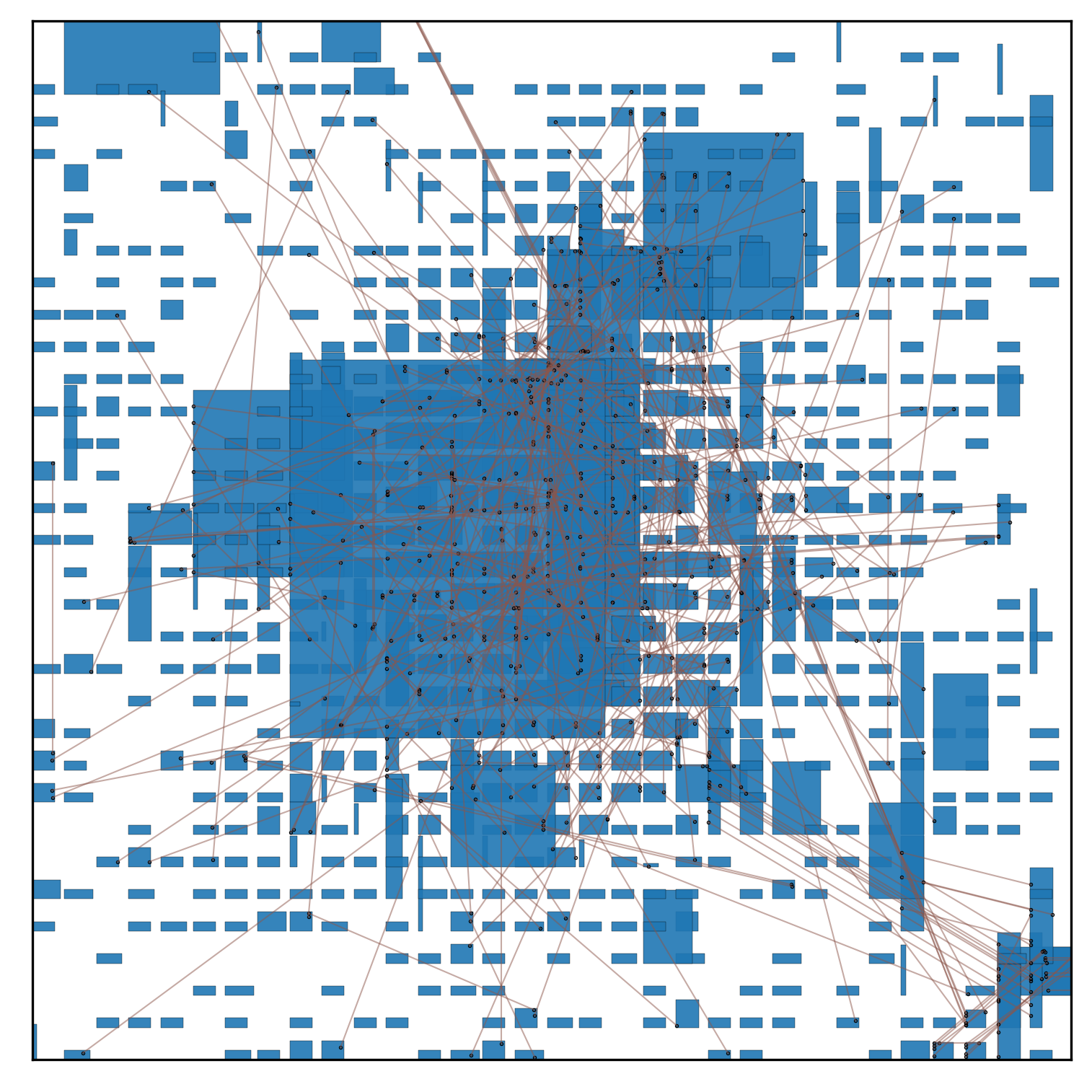}} \hfill
    \subfloat[Ours]{\includegraphics[width=0.48\linewidth]{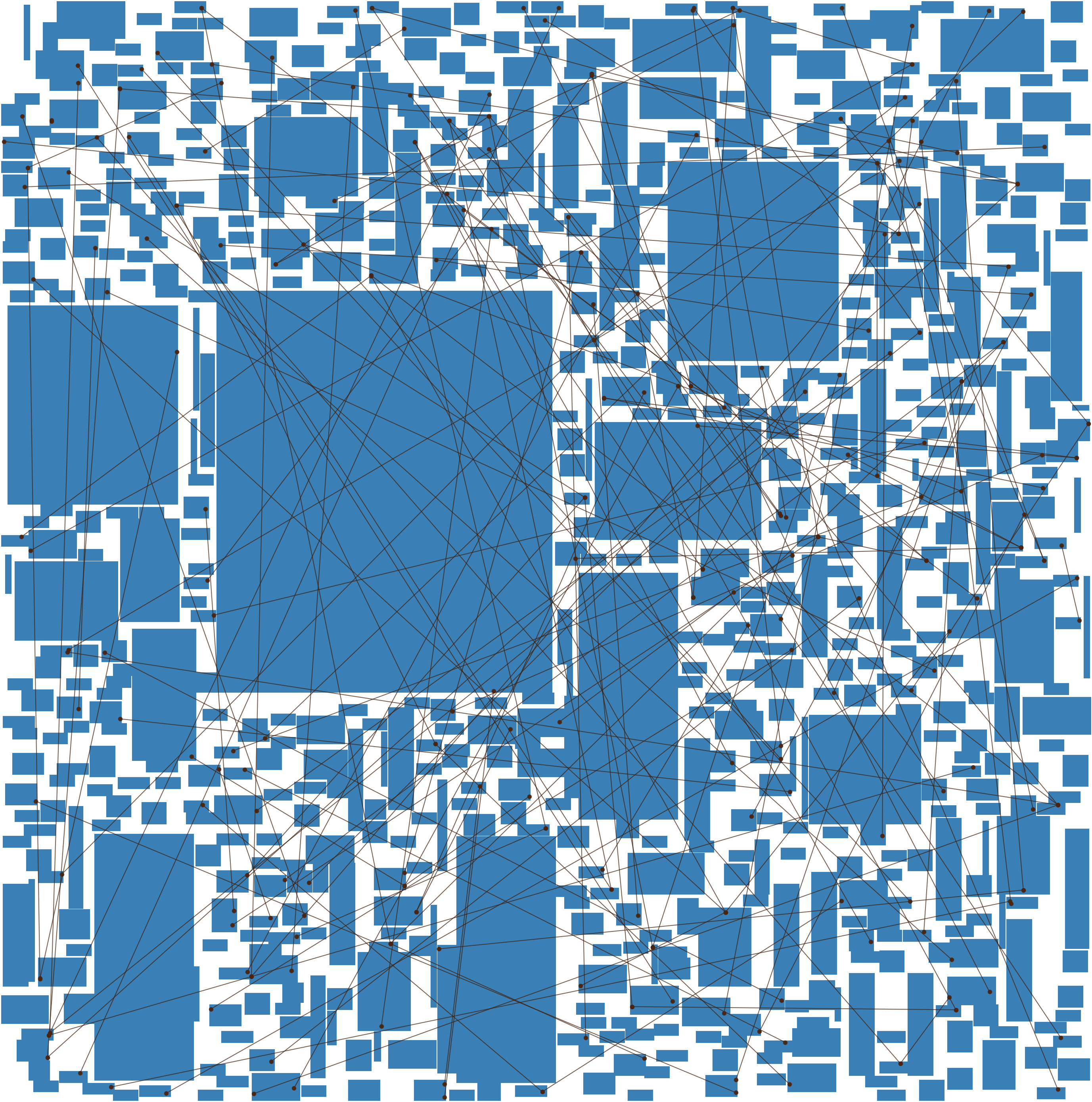}}
    \caption{Visualization of placement results on adaptec3 benchmark: (a) GraphPlace, (b) DeepPR (c) DREAMPlace, and (d) Ours. \modelname produces more compact placements with shorter wirelength compared to existing approaches.}
    \label{fig:visualization}
\end{figure}

Furthermore, the generated layouts in Fig.~\ref{fig:visualization} illustrate high structural regularity, suggesting that the energy-guided sampling implicitly balances wirelength minimization with density constraints without requiring manual tuning.

\begin{table*}[t]
\caption{Comparison of Macro-Only HPWL ($\times 10^5$) on ISPD Benchmarks .
\textbf{Bold} indicates the best result, and \underline{underlined} indicates the second-best.
``--'' denotes failure, timeout, or unreported results.}
\label{tab:ispd_macro_hpwl}
\centering
\resizebox{0.85\textwidth}{!}{
\begin{tabular}{lccccccc}
\toprule
\textbf{Benchmark} & \textbf{GraphPlace} & \textbf{DeepPR} & \textbf{DREAMPlace} & \textbf{MaskPlace} & \textbf{WireMask-BBO} & \textbf{ChiPFormer} & \textbf{\modelname (Ours)} \\
\midrule
adaptec1 
& 30.01 & 19.91 & 17.94 & 8.57 & \textbf{6.55} & \underline{6.82} & 12.15 \\
adaptec2 
& 351.71 & 203.51 & 135.32 & 77.70 & \textbf{53.80} & 64.20 & \underline{59.55} \\
adaptec3 
& 358.18 & 347.16 & 112.28 & 108.00 & \textbf{58.50} & 72.80 & \underline{60.10} \\
adaptec4 
& 151.42 & 311.86 & \textbf{37.77} & 91.90 & 61.50 & 84.90 & \underline{55.90} \\
bigblue1 
& 10.58 & 23.33 & \underline{2.50} & 3.11 & \textbf{2.05} & 3.15 & 5.80 \\
bigblue2 
& LG failed & LG failed & \underline{120.46} & Timeout & 185.33 & \textbf{84.51} & 138.50 \\
bigblue3 
& 357.48 & 430.48 & 104.05 & 84.00 & \underline{67.29} & 78.50 & \textbf{64.50} \\
bigblue4 
& 640.70 & 633.90 & -- & Timeout & 790.00 & \textbf{545.00} & \underline{615.07} \\
\bottomrule
\end{tabular}
}
\end{table*}

\begin{table}[t]
\caption{Comparison of Macro-Only HPWL ($\times 10^5$) on IBM Benchmarks.
\textbf{Bold} indicates the best result, and \underline{underlined} indicates the second-best.
}
\label{tab:ibm_macro_hpwl}
\centering
\resizebox{\columnwidth}{!}{
\begin{tabular}{lcccc}
\toprule
\textbf{Circuit} 
& \textbf{MaskPlace} 
& \textbf{WireMask-BBO} 
& \textbf{ChiPFormer} 
& \textbf{\modelname (Ours)} \\
\midrule
ibm01 & 4.30 & \textbf{2.78} & 3.88 & \underline{3.15} \\ 
ibm02 & 5.54 & \textbf{4.19} & 5.05 & \underline{4.45} \\ %
ibm03 & 3.31 & \textbf{3.30} & 3.74 & \underline{3.45} \\ %
ibm04 & 6.91 & \textbf{5.43} & 5.96 & \underline{5.60} \\ %
ibm06 & 0.93 & \underline{0.85} & 0.87 & \textbf{0.82} \\ 
ibm07 & 2.67 & 2.66 & \textbf{2.36} & \underline{2.48} \\ %
ibm08 & 20.60 & \textbf{19.20} & 19.90 & \underline{19.45} \\ %
ibm09 & 2.45 & \textbf{1.76} & \underline{1.77} & 1.95 \\     %
ibm10 & 23.80 & \textbf{18.20} & 18.25 & \underline{18.22} \\ %
ibm11 & 4.15 & 3.75 & \textbf{3.25} & 3.60 \\     
ibm12 & 14.90 & \underline{11.80} & 13.00 & \textbf{11.50} \\ 
ibm13 & 4.58 & 4.41 & \textbf{4.02} & 4.25 \\     
ibm14 & 8.43 & 9.80 & \textbf{7.44} & \underline{7.80} \\ %
ibm15 & 4.68 & 7.77 & \textbf{2.67} & 4.10 \\     %
ibm16 & 18.30 & \underline{14.80} & 15.50 & \textbf{14.50} \\ %
ibm17 & 16.80 & \textbf{12.20} & 13.70 & 13.10 \\    %
ibm18 & 5.98 & \textbf{3.44} & 4.19 & 3.90 \\     %

\midrule
\textbf{Avg.} & 8.72 & \textbf{7.43} & 7.79 & \underline{7.60} \\
\bottomrule
\end{tabular}
}
\end{table}

\subsubsection{Performance on industrial benchmark}
    

\begin{table}[t]
\caption{Comparison of Macro-Only HPWL ($\times 10^5$) on TILOS Benchmarks.
``Failed'' indicates the method failed to converge or generate a legal solution within 3 hours.
\textbf{Bold} indicates the best result.}
\label{tab:soc_macro_hpwl}
\centering
\resizebox{\columnwidth}{!}{
\begin{tabular}{lccccc}
\toprule
\textbf{Benchmark}
& \textbf{DREAMPlace}
& \textbf{MaskPlace}
& \textbf{WireMaskBBO}
& \textbf{ChiPFormer}
& \textbf{\modelname (Ours)} \\
\midrule
ariane133   & 15.20  & 16.85  & 17.10 & 16.50 & \textbf{12.71} \\
ariane136   & 18.10  & 19.42  & --    & --    & \textbf{15.44} \\
mempool     & 0.12   & --     & Failed & --    & \textbf{0.09} \\
nvdla       & 32.40  & Failed & Failed & Failed & \textbf{25.71} \\
\bottomrule
\end{tabular}
}
\end{table}

Table~\ref{tab:soc_macro_hpwl} reports the evaluation results on modern RISC-V (\texttt{ariane}, \texttt{mempool}) and deep learning accelerator (\texttt{nvdla}) designs. Unlike the synthetic ISPD benchmarks, these circuits represent real-world physical design challenges characterized by strict floorplanning constraints and the presence of dense standard cell clusters interleaved with memory blocks.

Although the macro count in these instances is moderate compared to the \texttt{bigblue} series, the primary challenge lies in generalization and legalization. Learning-based methods trained on legacy benchmarks often fail to adapt to the unseen topological structures of modern SoCs. As shown in Table~\ref{tab:soc_macro_hpwl}, baselines such as WireMask-BBO and ChiPFormer failed to generate legal solutions (or timed out) for \texttt{nvdla} and \texttt{mempool} due to their inability to satisfy the complex boundary conditions inherent in tiled architectures.
In contrast, \modelname demonstrates robust zero-shot generalization. Without any fine-tuning on these specific architectures, our model successfully generated legal placements with competitive wirelengths. The images illustrating ours placement results with industrial benchmarks are included in Figure \ref{fig:ispd05_final_results}.

\subsection{Mixed-Size Placement Results}
\subsubsection{Performance on academic benchmark}

Table \ref{tab:mixed_size_ispd05_final} presents the HPWL results of the hierarchical flow, where \modelname positions macros and DREAMPlace handles the subsequent standard cell placement. On large-scale benchmarks (\texttt{bigblue3}, \texttt{bigblue4}), \modelname achieves the lowest HPWL of $27.01 \times 10^7$ and $58.90 \times 10^7$ respectively. This performance indicates that the simultaneous diffusion process generates macro layouts that preserve contiguous whitespace, reducing fragmentation compared to sequential methods. Consequently, the downstream analytical placer can optimize standard cell distributions with fewer obstructions from sub-optimal macro blockages.Conversely, on smaller designs such as \texttt{adaptec1}, analytical (DREAMPlace) and search-based (WireMask-BBO) baselines perform marginally better. This reflects the algorithmic trade-off between global structure and local precision: while \modelname effectively captures the global connectivity basin for complex topologies, gradient-based solvers offer finer local granularity for simpler, less congested search spaces.

\begin{table*}[t]
\caption{Comparison of HPWL ($\times 10^7$) for mixed-size placement on ISPD05 benchmarks.
DREAMPlace results are based on the official v4.1.0 repository \protect\footnotemark.
\textbf{Bold} indicates the best result, and \underline{underlined} indicates the second-best.}
\label{tab:mixed_size_ispd05_final}
\centering
\resizebox{0.75\textwidth}{!}{
\begin{tabular}{lcccccccc}
\toprule
\multirow{2}{*}{\textbf{Method}}
& \multicolumn{8}{c}{\textbf{ISPD05}} \\
\cmidrule(lr){2-9}
& adaptec1 & adaptec2 & adaptec3 & adaptec4
& bigblue1 & bigblue2 & bigblue3 & bigblue4 \\
\midrule

MaskPlace + DP
& 7.93 & 9.95 & 22.97 & 22.97
& 9.43 & 14.13 & 37.29 & 106.18 \\

DREAMPlace 
& 6.82 & 8.63 & 14.40 & \textbf{14.08}
& \textbf{8.20} & 9.81 & 28.88 & 61.00 \\

WireMask-BBO + DP
& \textbf{6.45} & 8.90 & 13.90 & 14.50
& 8.85 & \underline{9.50} & 35.80 & 82.40 \\

ChiPFormer + DP
& \underline{6.55} & \textbf{7.36} & \textbf{13.07} & \underline{14.20}
& \underline{8.48} & \textbf{9.20} & \underline{27.85} & \underline{60.50} \\

\midrule
\textbf{\modelname (Ours)}
& 12.50
& \underline{7.91}
& \underline{13.42}
& 14.35
& 8.76
& 14.14
& \textbf{27.01}
& \textbf{58.90} \\
\bottomrule
\end{tabular}
}
\end{table*}

\footnotetext{DREAMPlace authors update the reference results on GitHub:
\url{https://github.com/limbo018/DREAMPlace}}

\begin{figure*}[t]
    \centering
    \subfloat[adaptec1]{
        \includegraphics[width=0.23\textwidth]{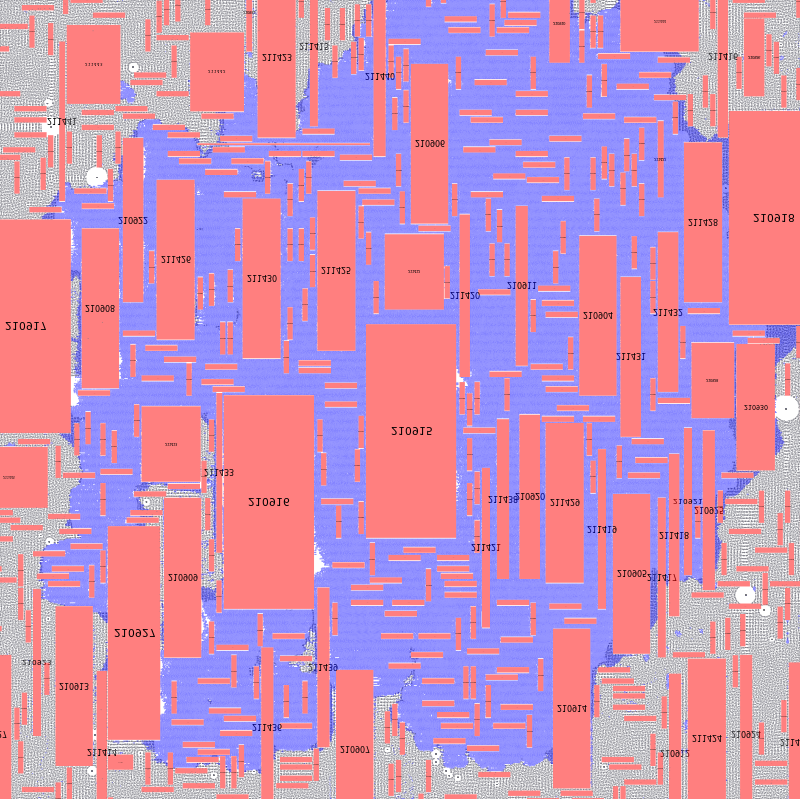}
        \label{fig:adaptec1}
    }
    \hfill
    \subfloat[adaptec2]{
        \includegraphics[width=0.23\textwidth]{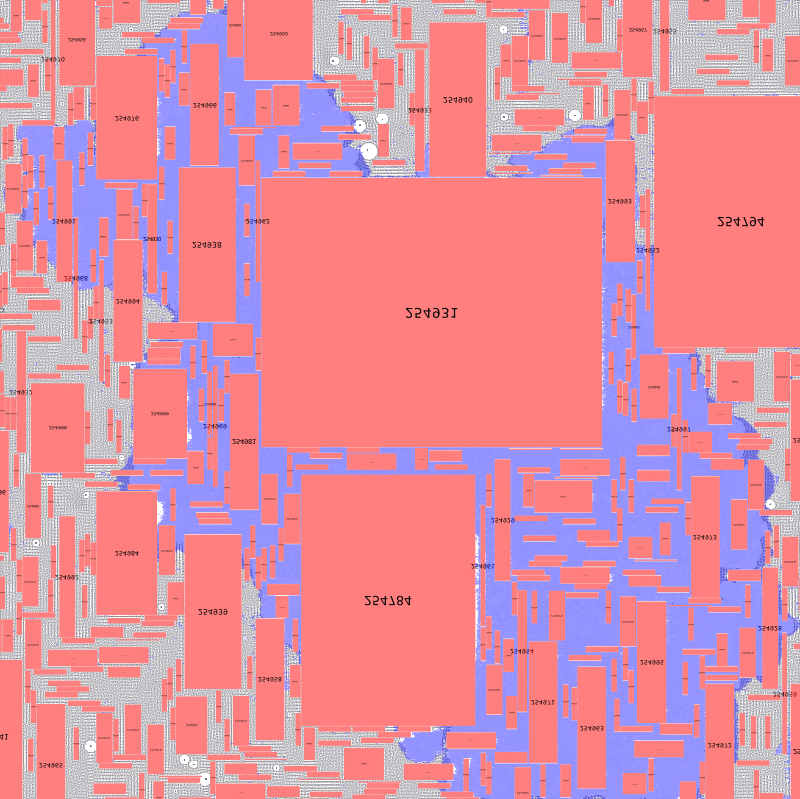}
        \label{fig:adaptec2}
    }
    \hfill
    \subfloat[adaptec3]{
        \includegraphics[width=0.23\textwidth]{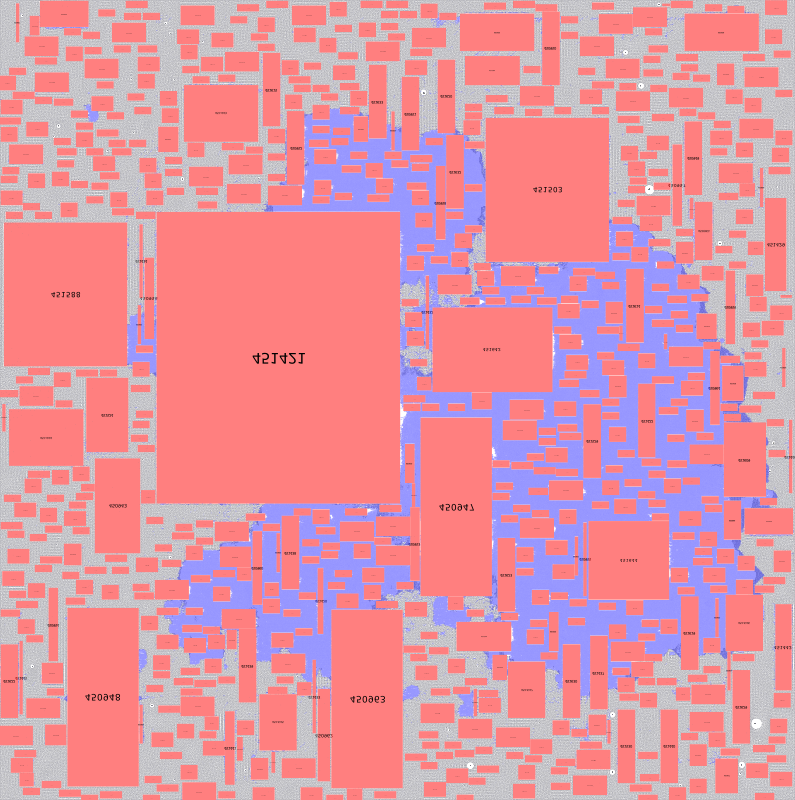}
        \label{fig:adaptec3}
    }
    \hfill
    \subfloat[adaptec4]{
        \includegraphics[width=0.23\textwidth]{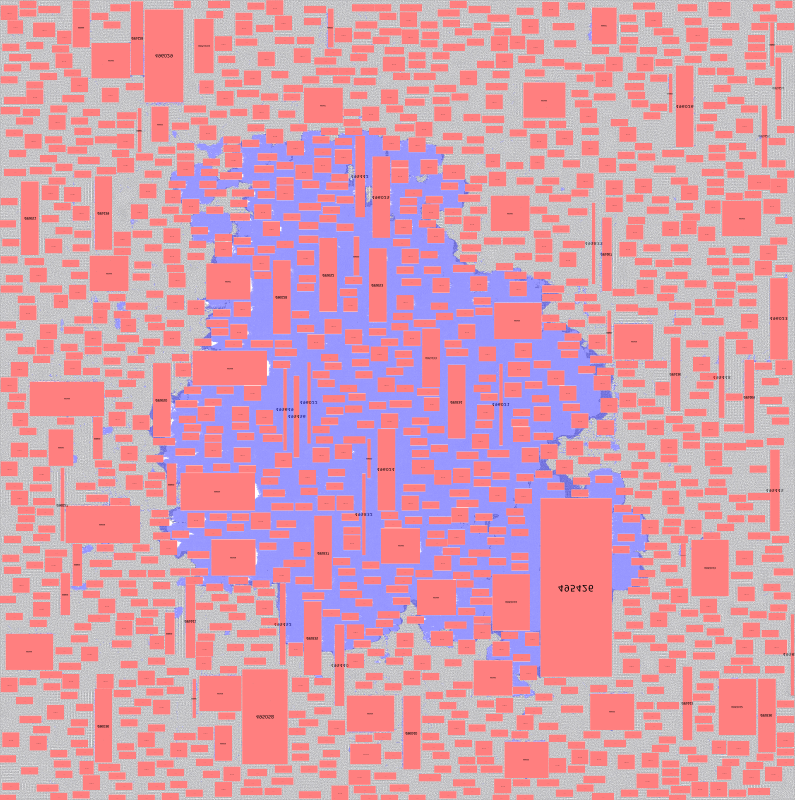}
        \label{fig:adaptec4}
    }
    
    \vspace{1ex} 
    
    \subfloat[bigblue1]{
        \includegraphics[width=0.23\textwidth]{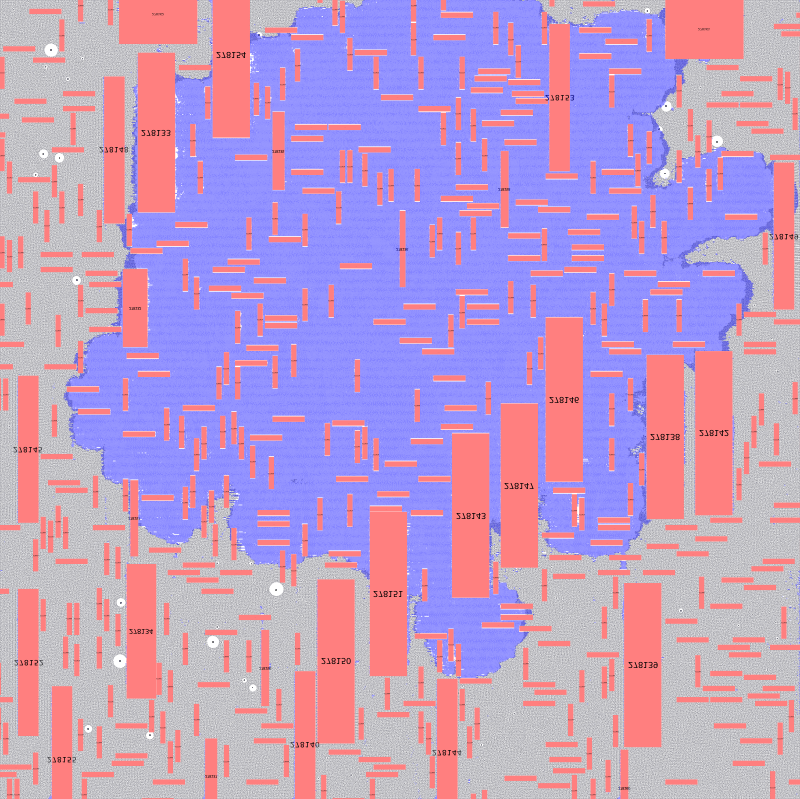}
        \label{fig:bigblue1}
    }
    \hfill
    \subfloat[bigblue2]{
        \includegraphics[width=0.23\textwidth]{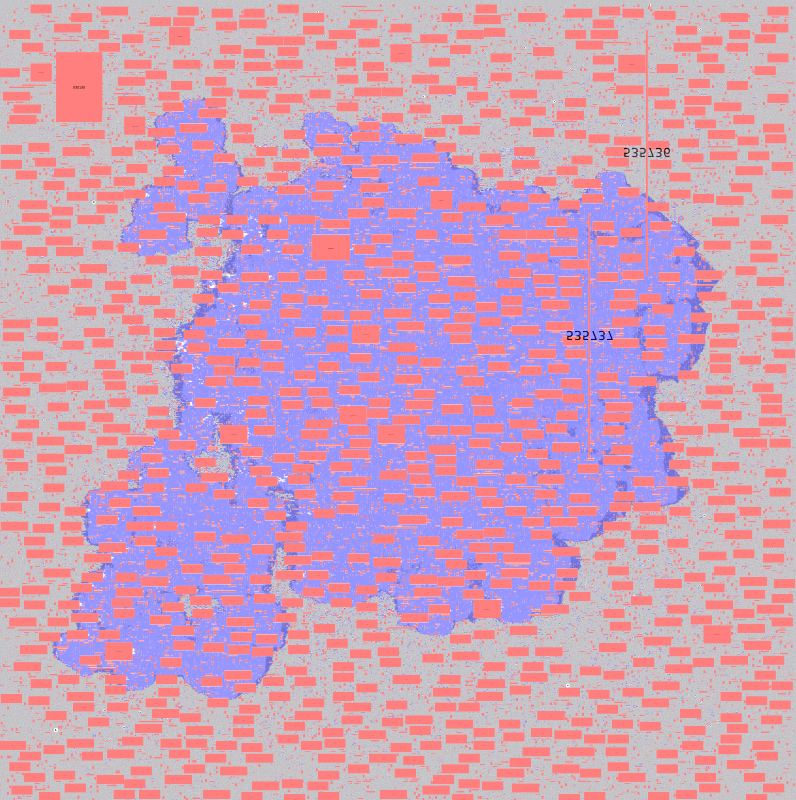}
        \label{fig:bigblue2}
    }
    \hfill
    \subfloat[bigblue3]{
        \includegraphics[width=0.23\textwidth]{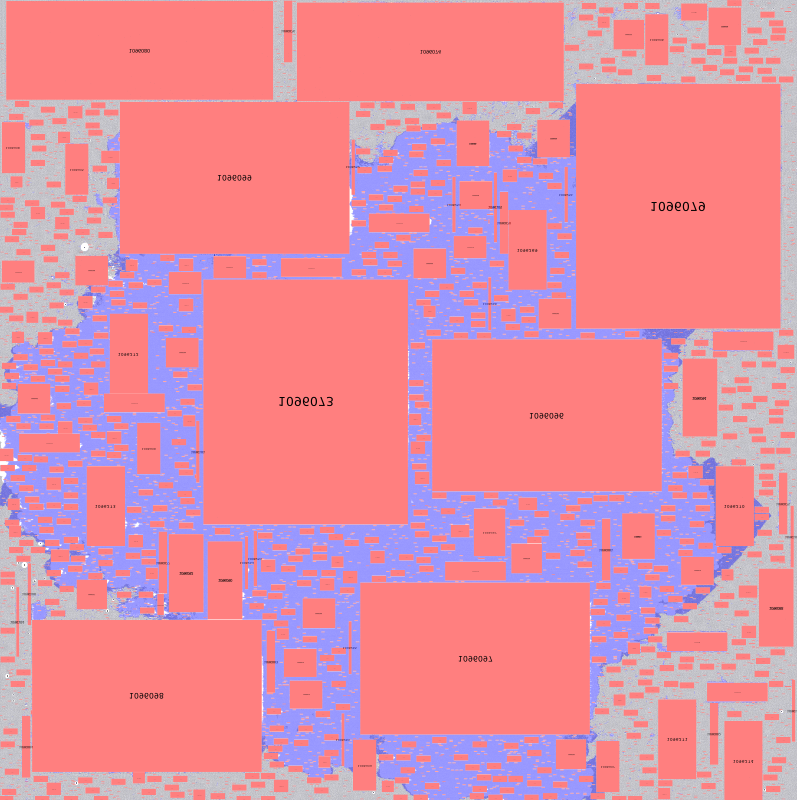}
        \label{fig:bigblue3}
    }
    \hfill
    \subfloat[bigblue4]{
        \includegraphics[width=0.23\textwidth]{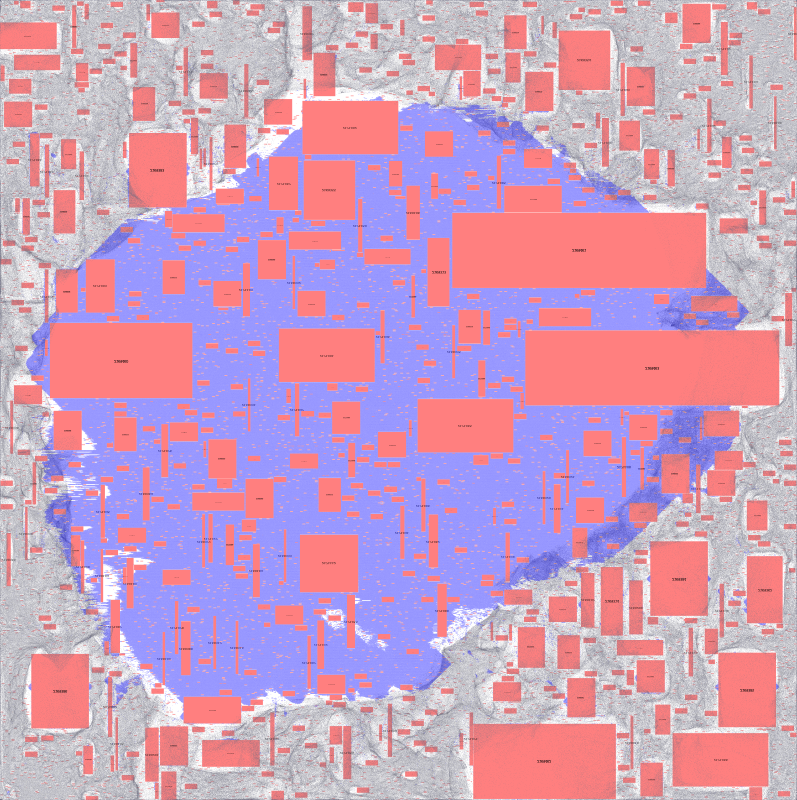}
        \label{fig:bigblue4}
    }
    \hfill
    \subfloat[ariane133]{
        \includegraphics[width=0.23\textwidth]{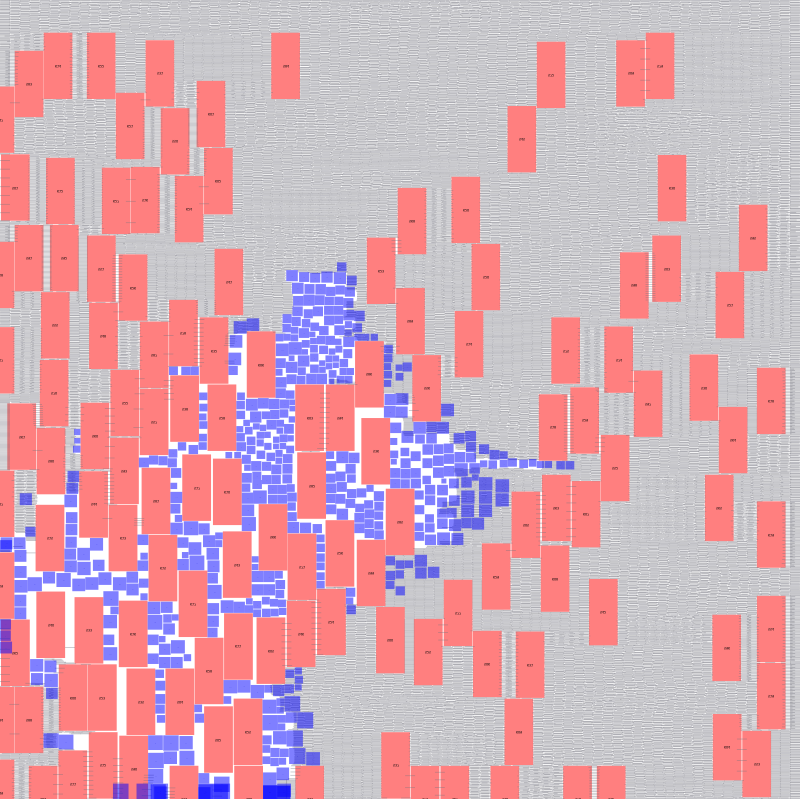}
        \label{fig:ariane133}
    }
    \hfill
    \subfloat[ariane136]{
        \includegraphics[width=0.23\textwidth]{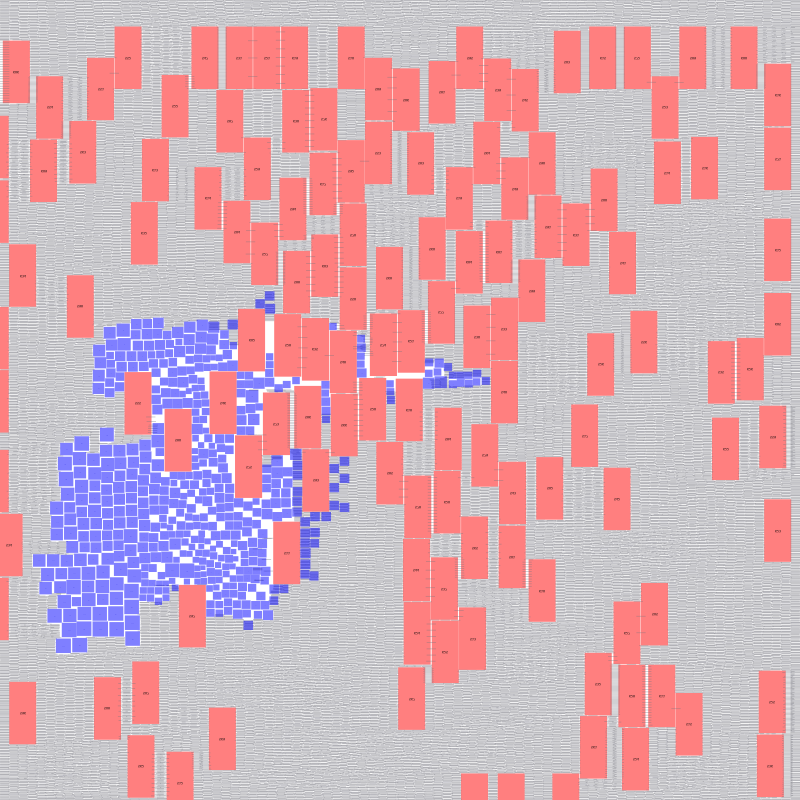}
        \label{fig:ariane136}
    }
    \hfill
    \subfloat[mempool-tile]{
        \includegraphics[width=0.23\textwidth]{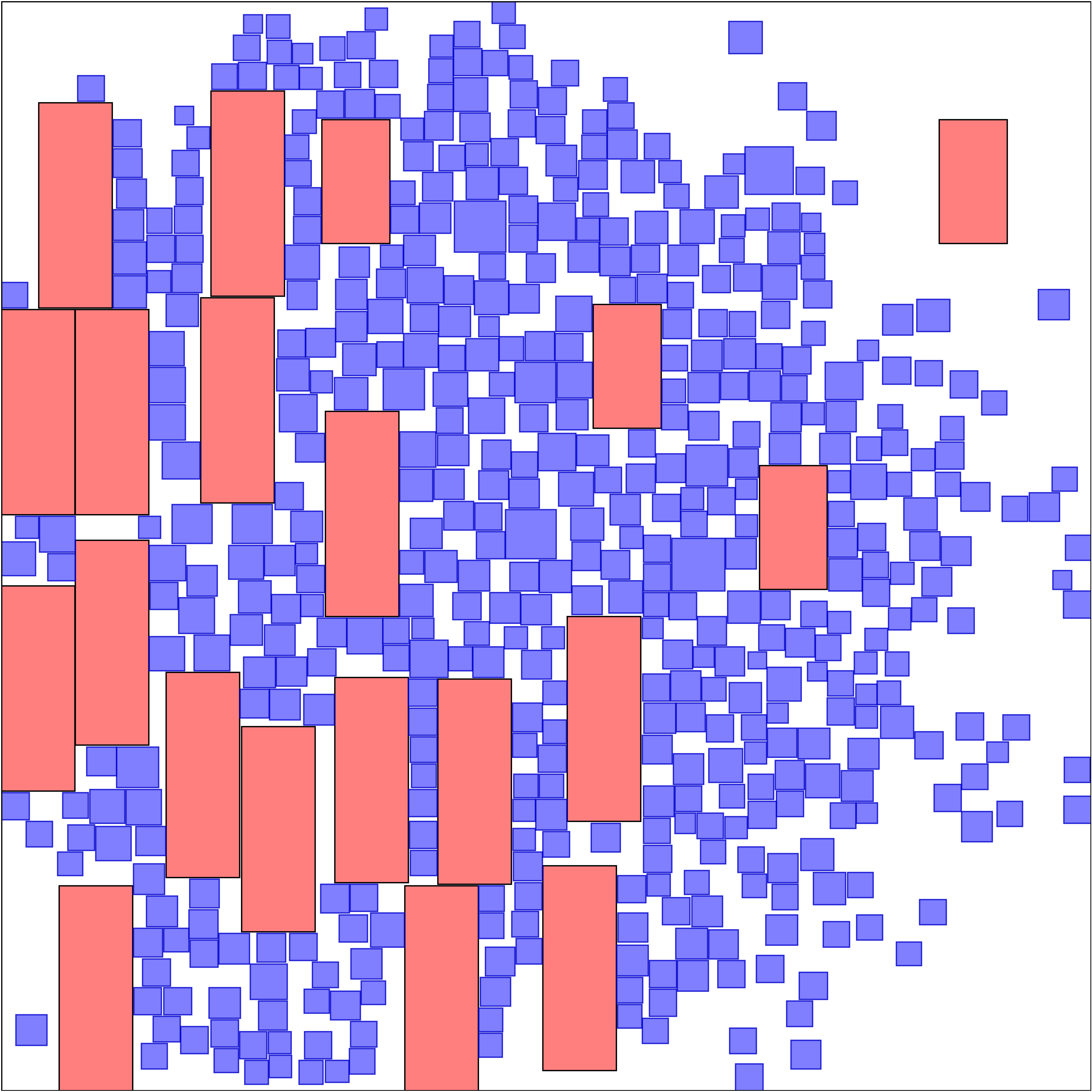}
        \label{fig:ariane136}
    }
    \hfill
    \subfloat[nvdla]{
        \includegraphics[width=0.23\textwidth]{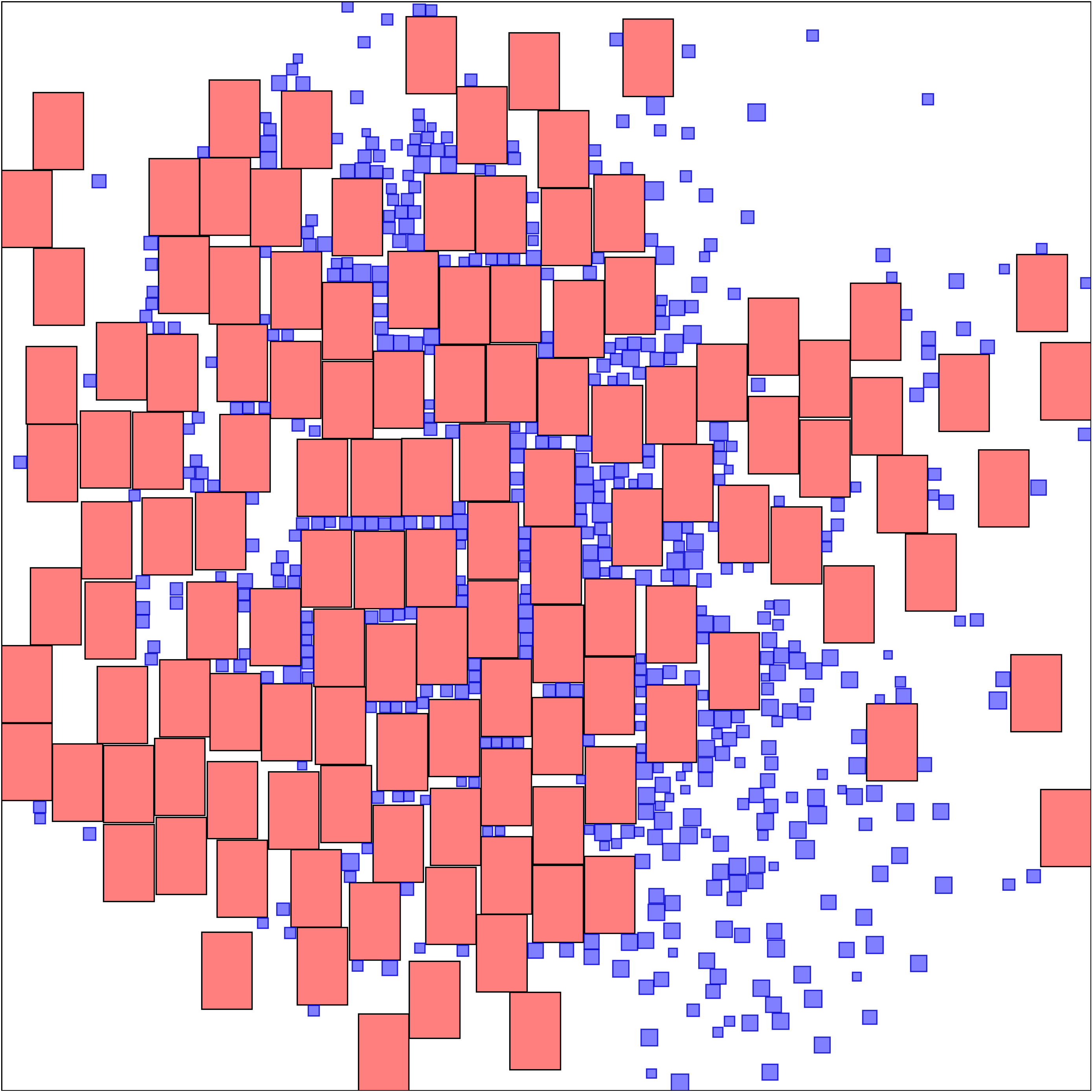}
        \label{fig:ariane136}
    }      
    \caption{Mixed-size placement results. (a)--(h): ISPD'05 benchmarks with macro placement from DiffPlace followed by DREAMPlace cell placement. (i)--(j): Ariane SoC designs with 133/136 SRAM macros. (k)--(l): Mempool-tile and NVDLA industrial designs showing SRAM macros (red) and clustered standard cells (blue) \textbf{simultaneously placed} by \modelname not using DREAMPlace.}
   \label{fig:ispd05_final_results}
\end{figure*}

\noindent \textbf{For mixed-size placement}, which involves positioning both macros and standard cells across the entire chip canvas, our methodology follows a three-stage hierarchical approach. This methodology is consistent with established practices in previous research~\cite{DBLP:conf/nips/LaiM022, DBLP:conf/nips/Shi0L023, DBLP:journals/corr/abs-2004-10746, DBLP:conf/nips/ChengY21}, while differing from the approach in~\cite{DBLP:conf/icml/LaiLTWH023}, where macro-displacement during standard cell placement obscures the direct impact of the initial macro-placement strategy. 
\noindent In the initial stage, we utilize our diffusion-based framework to perform Global Placement (GP) for all macros simultaneously. 
\noindent In the second stage, we implement a legalization (LG) step which presented in Algorithm~\ref{alg:inference_hdlp} to ensure macro positions conform to valid coordinates. 
\noindent The final stage preserves these fixed macro locations and utilizes DREAMPlace \cite{DBLP:journals/tcad/LinJGLDRKP21} as the dedicated standard cell placer. By inheriting initial coordinates from macro clusters and performing detailed placement optimization, we achieve a complete and optimized layout. The final results for the \texttt{adaptec}, \texttt{bigblue} series and industrial benchmarks are visualized in Figure \ref{fig:ispd05_final_results}.

Table~\ref{tab:mixed_size_ibm_final} illustrates the generalization of \modelname across the IBM benchmark. On smaller instances, search-based and analytical baselines retain a marginal advantage, utilizing exhaustive local refinement to fully exploit the constrained search spaces. However, as topological dimensions scale, our method establishes superior global optimization, effectively surpassing traditional baselines in high-complexity regimes. This crossover confirms that while heuristic solvers risk stagnation in local optima when facing explosive solution spaces, our diffusion-based energy guidance successfully captures global structural dependencies, ensuring robust scalability where traditional methods falter.

\begin{table*}[t]
\caption{Comparison of HPWL ($\times 10^6$) for mixed-size placement on IBM benchmarks.
\textbf{Bold} indicates the best result, and \underline{underlined} indicates the second-best.}
\label{tab:mixed_size_ibm_final}
\centering
\resizebox{\textwidth}{!}{
\begin{tabular}{lcccccccccccccccccc}
\toprule
\textbf{Benchmark} 
& \textbf{ibm01} & \textbf{ibm02} & \textbf{ibm03} & \textbf{ibm04} & \textbf{ibm05}
& \textbf{ibm06} & \textbf{ibm07} & \textbf{ibm08} & \textbf{ibm09} & \textbf{ibm10}
& \textbf{ibm11} & \textbf{ibm12} & \textbf{ibm13} & \textbf{ibm14} & \textbf{ibm15}
& \textbf{ibm16} & \textbf{ibm17} & \textbf{ibm18} \\
\midrule
MaskPlace
& 3.33 & 7.30 & 10.10 & 10.40 & 7.67
& 7.62 & 13.30 & 15.50 & 16.20 & 46.80
& \underline{23.50} & 46.10 & 28.20 & 45.40 & 53.40
& 65.90 & 72.90 & 42.20 \\

WireMask-BBO
& \textbf{2.15} & \textbf{4.35} & \textbf{9.81} & 9.65 & \underline{7.65}
& 8.41 & 13.00 & 15.90 & 15.40 & 45.20
& 24.60 & Failed & 28.00 & 48.20 & Failed
& 63.20 & 69.70 & 41.60 \\

ChiPFormer
& 3.35 & 6.24 & 10.90 & 10.10 & 7.67
& 7.76 & 13.40 & 15.70 & 16.90 & 45.40
& 23.60 & 48.80 & 28.40 & 46.50 & 55.80
& 67.30 & 71.40 & 41.10 \\

DREAMPlace
& \underline{2.23} & 5.79 & \underline{10.40} & \textbf{9.13} & \textbf{7.60}
& \underline{6.15} & \textbf{11.10} & \textbf{12.30} & \underline{12.80} & \underline{44.80}
& \textbf{16.60} & \underline{31.00} & \underline{23.20} & \textbf{31.30} & \underline{51.30}
& \underline{53.00} & \textbf{57.90} & \textbf{37.60} \\

\textbf{\modelname (Ours)}
& 2.35 & \underline{4.48} & 10.65 & \underline{9.25} & 7.82
& \textbf{6.08} & \underline{11.35} & 12.55 & \textbf{12.55} & \textbf{43.90}
& 26.90 & \textbf{30.45} & \textbf{22.85} & \underline{31.60} & \textbf{49.80}
& \textbf{51.90} & 58.60 & \underline{38.10} \\
\bottomrule
\end{tabular}
}
\end{table*}

\subsubsection{Performance on industrial benchmark}

Table~\ref{tab:soc_mixed_hpwl} reports the evaluation results on modern RISC-V (\texttt{ariane}, \texttt{mempool}) and deep learning accelerator (\texttt{nvdla}) designs. The final results for the  TILOS benchmarks are visualized in Figure \ref{fig:ispd05_final_results}.

\begin{table}[t]
\caption{Comparison of Mixed-Size HPWL ($\times 10^5$) on TILOS Benchmarks.}
\label{tab:soc_mixed_hpwl}
\centering
\resizebox{\columnwidth}{!}{
\begin{tabular}{lccccc}
\toprule
\textbf{Benchmark}
& \textbf{DREAMPlace}
& \textbf{MaskPlace}
& \textbf{WireMaskBBO}
& \textbf{ChiPFormer}
& \textbf{\modelname (Ours)} \\
\midrule
ariane133   & \textbf{183.11} & 212.54 & 215.00 & 213.43 & \underline{197.35} \\
ariane136   & \textbf{218.43} & --     & --     & -- & \underline{223.08} \\
mempool     & 108.55 & --     & --     & -- & \textbf{102.10} \\
nvdla       & 310.51 & Failed & Failed & Failed & \textbf{290.41} \\
\bottomrule
\end{tabular}
}
\end{table} 

\subsection{Congestion and Overlap }

Tables \ref{tab:rudy_ibm_final} and \ref{tab:rudy_ispd_final} present the routing congestion evaluation using the RUDY estimator. The routing congestion evaluation reveals the intrinsic trade-off between aggressive wirelength minimization and placement routability. While analytical solvers inherently excel in density smoothing on standard benchmarks due to their continuous objective functions, \modelname demonstrates a superior ability to navigate the multi objective optimization compared to other learning-based approaches. Unlike search-based heuristics or sequential reinforcement learning methods that often induce localized hotspots during iterative refinement, our global diffusion mechanism identifies a balanced manifold that minimizes wirelength without compromising placement feasibility. This architectural advantage allows \modelname to distribute logic density more evenly, mitigating severe routing bottlenecks even on high-complexity circuit topologies where traditional local search algorithms falter.


\begin{table*}[t]
\caption{Comparison of RUDY congestion estimator on IBM benchmarks.
\textbf{Bold} indicates the best result, and \underline{underlined} indicates the second-best.}
\label{tab:rudy_ibm_final}
\centering
\resizebox{\textwidth}{!}{
\begin{tabular}{lccccccccccccccccc}
\toprule
\textbf{Benchmark}
& \textbf{ibm01} & \textbf{ibm02} & \textbf{ibm03} & \textbf{ibm04}
& \textbf{ibm06} & \textbf{ibm07} & \textbf{ibm08} & \textbf{ibm09}
& \textbf{ibm10} & \textbf{ibm11} & \textbf{ibm12} & \textbf{ibm13}
& \textbf{ibm14} & \textbf{ibm15} & \textbf{ibm16} & \textbf{ibm17}
& \textbf{ibm18} \\
\midrule
MaskPlace
& 291.0 & \textbf{225.0} & \underline{178.0} & \textbf{452.0}
& 78.5 & \textbf{151.0} & 1235.0 & 125.0
& 483.0 & 184.0 & 395.0 & \textbf{160.0}
& 381.0 & \textbf{160.0} & 571.0 & 535.0
& 274.0 \\

WireMask-BBO
& 282.0 & 245.0 & 182.0 & 480.0
& 79.0 & 166.0 & \underline{1192.0} & \underline{116.0}
& \textbf{461.0} & 181.0 & \underline{215.0} & 205.0
& \textbf{372.0} & 175.0 & \underline{494.0} & \textbf{461.0}
& \textbf{218.0} \\

ChiPFormer
& \underline{263.0} & \underline{232.0} & 181.0 & 493.0
& \underline{75.8} & 163.0 & 1265.0 & \textbf{113.0}
& 469.0 & \underline{175.0} & 354.0 & 179.0
& \underline{375.0} & 176.0 & 531.0 & 485.0
& 232.0 \\

\textbf{\modelname (Ours)}
& \textbf{254.0} & 239.0 & \textbf{173.0} & \underline{464.0}
& \textbf{71.0} & \underline{156.0} & \textbf{1175.0} & 119.0
& \underline{467.0} & \textbf{169.0} & \textbf{208.0} & \underline{172.0}
& 383.0 & \underline{166.0} & \textbf{482.0} & \underline{472.0}
& \underline{222.0} \\
\bottomrule
\end{tabular}
}
\end{table*}



\begin{table*}[t]
\caption{Comparison of RUDY congestion estimator on ISPD2005 benchmarks.
\textbf{Bold} indicates the best result, and \underline{underlined} indicates the second-best.}
\label{tab:rudy_ispd_final}
\centering
\resizebox{0.85\textwidth}{!}{
\begin{tabular}{lcccccccc}
\toprule
\textbf{Method}
& adaptec1 & adaptec2 & adaptec3 & adaptec4
& bigblue1 & bigblue2 & bigblue3 & bigblue4 \\
\midrule
MaskPlace
& 315.0 & 1072.0 & 994.0 & 942.0
& 99.0 & Timeout & 974.0 & Timeout \\

WireMask-BBO
& \textbf{136.0} & 1080.0 & \underline{214.0} & 796.0
& \underline{24.5} & 1920.0 & 952.0 & 6295.0 \\

ChiPFormer
& \underline{142.0} & 1175.0 & 674.0 & 782.0
& \textbf{21.0} & \underline{505.0} & 959.0 & 2432.0 \\

DREAMPlace
& 319.0 & \textbf{255.0} & \textbf{212.0} & \textbf{478.0}
& 314.0 & \textbf{255.0} & \textbf{216.0} & \textbf{482.0} \\

\modelname (Ours)
& 148.0 & \underline{298.0} & 248.0 & \underline{523.0}
& 29.0 & 583.0 & \underline{253.0} & \underline{512.0} \\
\bottomrule
\end{tabular}
}
\end{table*}

Unlike legacy academic benchmarks which model placement on largely unconstrained canvases, TILOS benchmark introduces the rigorous complexity of modern industrial SoCs. These designs are characterized by strict floorplanning constraints, including defined fence regions for specific voltage domains and hierarchical tiling structures typical of multi-core RISC-V and AI accelerator architectures. Consequently, these benchmarks test the placer's ability to satisfy hard geometric boundary conditions and logical isolation requirements, moving beyond simple wirelength minimization to address the feasibility challenges of real-world physical design.

\begin{table*}[t]
\caption{Comparison of placement quality (Congestion \& Overlap) on TILOS Benchmarks.}
\label{tab:constraint_comparison}
\centering
\resizebox{0.85\textwidth}{!}{
\begin{tabular}{lcccccccc}
\toprule
\multirow{2}{*}{Benchmark}
& \multicolumn{2}{c}{DREAMPlace}
& \multicolumn{2}{c}{MaskPlace}
& \multicolumn{2}{c}{ChiPFormer}
& \multicolumn{2}{c}{\modelname\ (Ours)} \\
\cmidrule(lr){2-3}
\cmidrule(lr){4-5}
\cmidrule(lr){6-7}
\cmidrule(lr){8-9}
& Cong. & Overlap (\%)
& Cong. & Overlap (\%)
& Cong. & Overlap (\%)
& Cong. & Overlap (\%) \\
\midrule
ariane133     & 1.25 & 0.05 & 1.42 & 0.00 & 1.38 & 0.00 & \textbf{1.09} & \textbf{0.00} \\
ariane136     & 1.15 & 0.08 & --   & --   & --   & --   & \textbf{0.87} & \textbf{0.00} \\
mempool       & 0.35 & 0.02 & --   & --   & --   & --   & \textbf{0.26} & \textbf{0.00} \\
nvdla         & 18.50& 0.12 & --   & --   & --   & --   & \textbf{16.67}& \textbf{0.00} \\
\bottomrule
\end{tabular}
}
\end{table*}

\subsection{Zero-shot transfer Capabilities}
\subsubsection{Zeroshot Learning with Modern Designs}

A bottleneck in reinforcement learning-based placement is the tendency to overfit policies to specific netlist topologies, hindering portability. \modelname overcomes this by learning a generalized diffusion trajectory derived from a diverse corpus of synthetic data. We pre-train the model on synthetic data generated with varying connectivity complexities and aspect ratios. This process forces the diffusion network to internalize fundamental physical placement rules, such as minimizing wirelength for highly connected clusters and managing whitespace fragmentation, rather than memorizing benchmark-specific artifacts. Additional illustration is provided in the Appendix. 

The efficacy of this approach is validated by the direct zero-shot transfer to the unseen TILOS Benchmarks Figure \ref{fig:ispd05_final_results}. Without any fine-tuning, \modelname successfully translates the macro-packing intuition learned from synthetic data to real-world designs, generating legal and optimized floorplans. 



\subsubsection{Runtime Analysis}

The runtime analysis across ISPD, IBM, and TILOS benchmarks in Table~\ref{tab:runtime_ispd}, \ref{tab:runtime_ibm}, \ref{tab:runtime_tilos} demonstrates an efficiency hierarchy dictated by the underlying algorithmic paradigms. While analytical solvers establish the speed lower bound through highly parallelized gradient descent, reinforcement learning and search-based baselines incur prohibitive computational costs due to the necessity of instance-specific training, often resulting in timeouts on large-scale designs. \modelname bridges this gap by leveraging a zero-shot inference mechanism; although the reverse diffusion sampling process inherently requires more compute cycles than a single gradient step that results in a moderate runtime increase over the analytical baseline, it effectively eliminates the iterative training overhead. Consequently, our approach achieves an acceleration of orders of magnitude compared to prior learning-based methods, validating that generative placement can be practically integrated into iterative physical design flows without the bottleneck of per-instance optimization.

\begin{table*}[t]
\caption{Comparison of runtime (minutes) for macro placement on the ISPD benchmark.}
\label{tab:runtime_ispd}
\centering
\resizebox{0.85\textwidth}{!}{
\begin{tabular}{lcccccccc}
\toprule
\textbf{Method}
& adaptec1 & adaptec2 & adaptec3 & adaptec4
& bigblue1 & bigblue2 & bigblue3 & bigblue4 \\
\midrule
MaskPlace
& 142.0 & 192.0 & 221.0 & 720.0
& 270.0 & Timeout & 651.0 & Timeout \\

WireMask-BBO
& 215.0 & 205.0 & 210.0 & 216.0
& 201.0 & 1402.0 & 230.0 & 601.0 \\

ChiPFormer
& 220.0 & 231.0 & 280.0 & 462.0
& 252.0 & 5210.0 & 490.0 & 1205.0 \\

DREAMPlace
& 0.46 & 0.67 & 0.94 & 0.96
& 0.43 & 3.22 & 2.34 & 3.91 \\

\modelname (Ours)
& 6.69 & 6.34 & 6.62 & 6.92
& 6.76 & 156.00 & 7.18 & 47.25 \\
\bottomrule
\end{tabular}
}
\end{table*}

\begin{table*}[t]
\caption{Component ablation study on \texttt{adaptec3} (ISPD05) and \texttt{nvdla} (TILOS).
We evaluate the contribution of key architectural modules to placement quality.}
\label{tab:ablation_dual}
\centering
\resizebox{0.65\textwidth}{!}{
\begin{tabular}{lcccc}
\toprule
\multirow{2}{*}{\textbf{Configuration}}
& \multicolumn{2}{c}{\textbf{adaptec3 (ISPD05)}}
& \multicolumn{2}{c}{\textbf{nvdla (TILOS)}} \\
\cmidrule(lr){2-3} \cmidrule(lr){4-5}
& HPWL ($\times 10^6$) & Overlap (\%)
& HPWL ($\times 10^6$) & Overlap (\%) \\
\midrule
\textbf{\modelname (Full)}
& \textbf{60.10} & \textbf{0.00}
& \textbf{45.50} & \textbf{0.00} \\
\midrule
w/o Global Context
& 72.45 & 0.00
& 58.20 & 0.05 \\
w/o Relative Pos. Emb.
& 98.60 & 1.25
& 85.40 & 2.10 \\
w/o Graph Topology
& 145.20 & 8.50
& 132.00 & 12.40 \\
Random Initialization
& 155.50 & 25.40
& 95.00 & 32.10 \\
\bottomrule
\end{tabular}
}
\end{table*}

\begin{table*}[t]
\caption{Comparison of runtime (minutes) for macro placement on the IBM benchmark.
Circuit \texttt{ibm05} is omitted as it contains no macros.}
\label{tab:runtime_ibm}
\centering
\resizebox{\textwidth}{!}{
\begin{tabular}{lccccccccccccccccc}
\toprule
\textbf{Method} 
& ibm01 & ibm02 & ibm03 & ibm04 & ibm06 & ibm07 & ibm08 & ibm09 & ibm10 & ibm11 & ibm12 & ibm13 & ibm14 & ibm15 & ibm16 & ibm17 & ibm18 \\
\midrule
MaskPlace 
& 151.0 & 162.0 & 120.0 & 66.0 & 36.0 & 60.0 & 72.0 & 53.0 & 512.0 & 81.0 & 385.0 & 91.0 & 390.0 & 85.0 & 104.0 & 482.0 & 63.0 \\

WireMask-BBO 
& 205.0 & 201.0 & 220.0 & 211.0 & 221.0 & 220.0 & 209.0 & 218.0 & 230.0 & 222.0 & 250.0 & 228.0 & 264.0 & 251.0 & 219.0 & 263.0 & 214.0 \\

ChiPFormer 
& 100.0 & 89.0 & 73.0 & 80.0 & 82.0 & 83.0 & 103.0 & 70.0 & 241.0 & 104.0 & 199.0 & 125.0 & 190.0 & 110.0 & 135.0 & 254.0 & 91.0 \\

EfficientPlace 
& 52.0 & 64.0 & 60.0 & 63.0 & 30.0 & 61.0 & 81.0 & 49.0 & 265.0 & 82.0 & 202.0 & 93.0 & 214.0 & 88.0 & 128.0 & 360.0 & 56.0 \\

DREAMPlace 
& 0.31 & 0.41 & 0.39 & 0.40 & 0.23 & 0.26 & 0.26 & 0.26 & 0.46 & 0.30 & 0.47 & 0.61 & 0.76 & 0.92 & 0.78 & 0.84 & 0.74 \\

\modelname (Ours) 
& 2.59 & 3.15 & 3.04 & 3.09 & 3.33 & 4.02 & 4.73 & 4.31 & 11.25 & 8.50 & 13.10 & 10.48 & 14.88 & 17.03 & 19.70 & 25.83 & 20.15 \\
\bottomrule
\end{tabular}
}
\end{table*}

\begin{table}[t]
\caption{Comparison of runtime (minutes) for macro placement on modern TILOS benchmarks.
``Timeout'' denotes runtime exceeding 3 hours.
\textbf{Bold} indicates the fastest method.}
\label{tab:runtime_tilos}
\centering
\resizebox{\columnwidth}{!}{
\begin{tabular}{lccccc}
\toprule
Circuit & MaskPlace & WireMaskBBO & ChiPFormer & DREAMPlace & \modelname \\
\midrule
ariane133 & 325.0 & 410.0 & 225.0 & \textbf{0.85} & 5.40 \\
ariane136 & 340.0 & Timeout   & 248.0 & \textbf{0.92} & 5.85 \\
mempool   & Timeout   & Timeout   & 415.0 & \textbf{4.15} & 18.50 \\
nvdla     & Timeout   & N/A   & Timeout & \textbf{3.60} & 15.20 \\
\bottomrule
\end{tabular}
}
\end{table}

\subsection{Ablation Studies and Analysis}
\subsubsection{Component Contribution Analysis}

We define the implementation changes for each configuration and analyze the corresponding performance shifts as follows:

\begin{itemize}
    \item \textbf{w/o Global Context:} This configuration removes the virtual supernode and gated broadcast mechanism, restricting the GNN to local neighborhood aggregation only. 

    \item \textbf{w/o Relative Pos. Emb.:} We replaced the layout-agnostic relative sinusoidal encodings ($\text{PE}(x_i - x_j)$) with standard absolute position embeddings.

    \item \textbf{w/o Graph Topology:} We disabled the message-passing operations between nodes, effectively reducing the architecture to a coordinate-based MLP.
\end{itemize}

Figure \ref{fig:training-loss} visualizes the training dynamics of the pre-training phase. The convergence curves demonstrate that the full \modelname architecture (blue) achieves the lowest and most stable loss. In contrast, removing key components, particularly the graph topology (red) and relative position embeddings (orange), leads to significantly slower convergence and higher final loss plateaus.

Table \ref{tab:ablation_dual} indicates that removing the global context module increases HPWL, suggesting that local message passing alone is limited in optimizing long-range macro dependencies. The substitution of relative embeddings with absolute coordinates leads to a performance decline on the unseen \texttt{nvdla} benchmark, implying that absolute positions tend to overfit training spatial distributions, whereas relative encoding facilitates generalization. Furthermore, eliminating the graph topology results in increased overlap and wirelength, showing that explicit connectivity modeling is necessary for resolving physical constraints.

\begin{figure}[t]
\centering
\begin{tikzpicture}
\begin{axis}[
    width=\linewidth,
    height=4.5cm,
    xlabel={Step},
    ylabel={Loss},
    xmin=0,
    ymin=0.7,
    ymax=1.05,
    grid=major,
    grid style={dotted,gray!30},
    tick label style={font=\footnotesize},
    label style={font=\footnotesize},
    line join=round,
    line cap=round,
    title={Pre-training},
    title style={font=\footnotesize\bfseries},
    no marks,
    unbounded coords=discard,
    legend style={font=\footnotesize, draw=none, fill=none},
    legend pos=north east,
]
\addplot[name path=full_u, draw=none]
table[col sep=tab, x index=0, y index=2] {data/loss_curve_pretraining_band.txt};
\addplot[name path=full_l, draw=none]
table[col sep=tab, x index=0, y index=1] {data/loss_curve_pretraining_band.txt};
\addplot[fill=blue!55, fill opacity=0.25, draw=none, forget plot]
fill between[of=full_u and full_l];

\addplot[draw=blue!85!black, line width=0.6pt]
table[col sep=tab, x index=0, y index=1] {data/loss_curve_pretraining_ema.txt};

\addplot[name path=gt_u, draw=none]
table[col sep=tab, x index=0, y index=2] {data/loss_curve_pretraining_wo_graph_topology_band.txt};
\addplot[name path=gt_l, draw=none]
table[col sep=tab, x index=0, y index=1] {data/loss_curve_pretraining_wo_graph_topology_band.txt};
\addplot[fill=red!60, fill opacity=0.18, draw=none, forget plot]
fill between[of=gt_u and gt_l];

\addplot[draw=red!75!black, line width=0.6pt]
table[col sep=tab, x index=0, y index=1] {data/loss_curve_pretraining_wo_graph_topology_ema.txt};

\addplot[name path=rp_u, draw=none]
table[col sep=tab, x index=0, y index=2] {data/loss_curve_pretraining_wo_relative_pos_emb_band.txt};
\addplot[name path=rp_l, draw=none]
table[col sep=tab, x index=0, y index=1] {data/loss_curve_pretraining_wo_relative_pos_emb_band.txt};
\addplot[fill=orange!70, fill opacity=0.18, draw=none, forget plot]
fill between[of=rp_u and rp_l];

\addplot[draw=orange!85!black, line width=0.6pt]
table[col sep=tab, x index=0, y index=1] {data/loss_curve_pretraining_wo_relative_pos_emb_ema.txt};

\addplot[name path=gc_u, draw=none]
table[col sep=tab, x index=0, y index=2] {data/loss_curve_pretraining_wo_global_context_band.txt};
\addplot[name path=gc_l, draw=none]
table[col sep=tab, x index=0, y index=1] {data/loss_curve_pretraining_wo_global_context_band.txt};
\addplot[fill=green!58, fill opacity=0.18, draw=none, forget plot]
fill between[of=gc_u and gc_l];

\addplot[draw=green!60!black, line width=0.6pt]
table[col sep=tab, x index=0, y index=1] {data/loss_curve_pretraining_wo_global_context_ema.txt};
\addlegendimage{line legend, draw=green!60!black, line width=0.6pt}

\end{axis}
\end{tikzpicture}

\vspace{4pt}
{\footnotesize Pre-training on large training dataset.}

\caption{
Training loss curves for the full DiffPlace model and three architecture ablations. Full (blue solid), w/o Graph Topology (red), w/o Relative Pos. Emb. (orange), w/o Global Context (green).
}
\label{fig:training-loss}
\end{figure}

\subsection{Limitations and Future Work}
Although \modelname demonstrates strong performance across benchmarks, we observe some limitations. Circuits with extremely sparse connectivity (fanout $<3$) occasionally show suboptimal results, as our GNN conditioning relies on rich connectivity patterns. Additionally, our current implementation focuses on placement quality metrics (HPWL, congestion, overlap), but does not explicitly optimize for timing closure, which remains critical for high-performance designs.

Future work will address timing-aware placement through the incorporation of delay models into our energy function, scaling to modern industrial circuit sizes (more than $10$M standard cells) and investigating hierarchical diffusion models for extremely large designs.

%% file: sections/conclusion.tex
\section{Conclusions}
\label{sec:conclusion}

In this work, we introduced \modelname, a novel diffusion-based generative framework for simultaneous VLSI chip placement. By modeling placement as a denoising diffusion process conditioned on netlist graphs and energy scores, \modelname eliminates the need for sequential module decisions and achieves fully parallel, constraint-aware optimization. Our design supports both macro-only and mixed-size placements, enforces zero-overlap through constraint satisfaction, and leverages a learnable energy-based guidance mechanism for improved placement quality.

Extensive experiments on the ISPD05 benchmark suite demonstrate that \modelname matches or outperforms prior methods across half-perimeter wirelength (HPWL) and constraint adherence while offering greater generalization and efficiency. Compared to analytical, reinforcement learning, and transformer-based approaches, our method delivers competitive placement quality with improved scalability and transferability.

In general, \modelname opens a new direction for integrating score-based generative modeling into VLSI design automation. Future work will extend the framework toward full placement-and-routing co-optimization, explore integration with commercial toolchains, and investigate broader applications of conditional diffusion in physical design tasks.

%% file: sections/appendix.tex
\section{Appendix}
\label{sec:Appendix}

\subsection{Visualization of Input Representations}
\label{app:data_visualization}

\begin{figure*}[t]
    \centering
    \setlength{\tabcolsep}{2pt}
    \begin{tabular}{ccc}
        \subfloat[Syn-Adaptec3]{
            \includegraphics[width=0.2\linewidth]{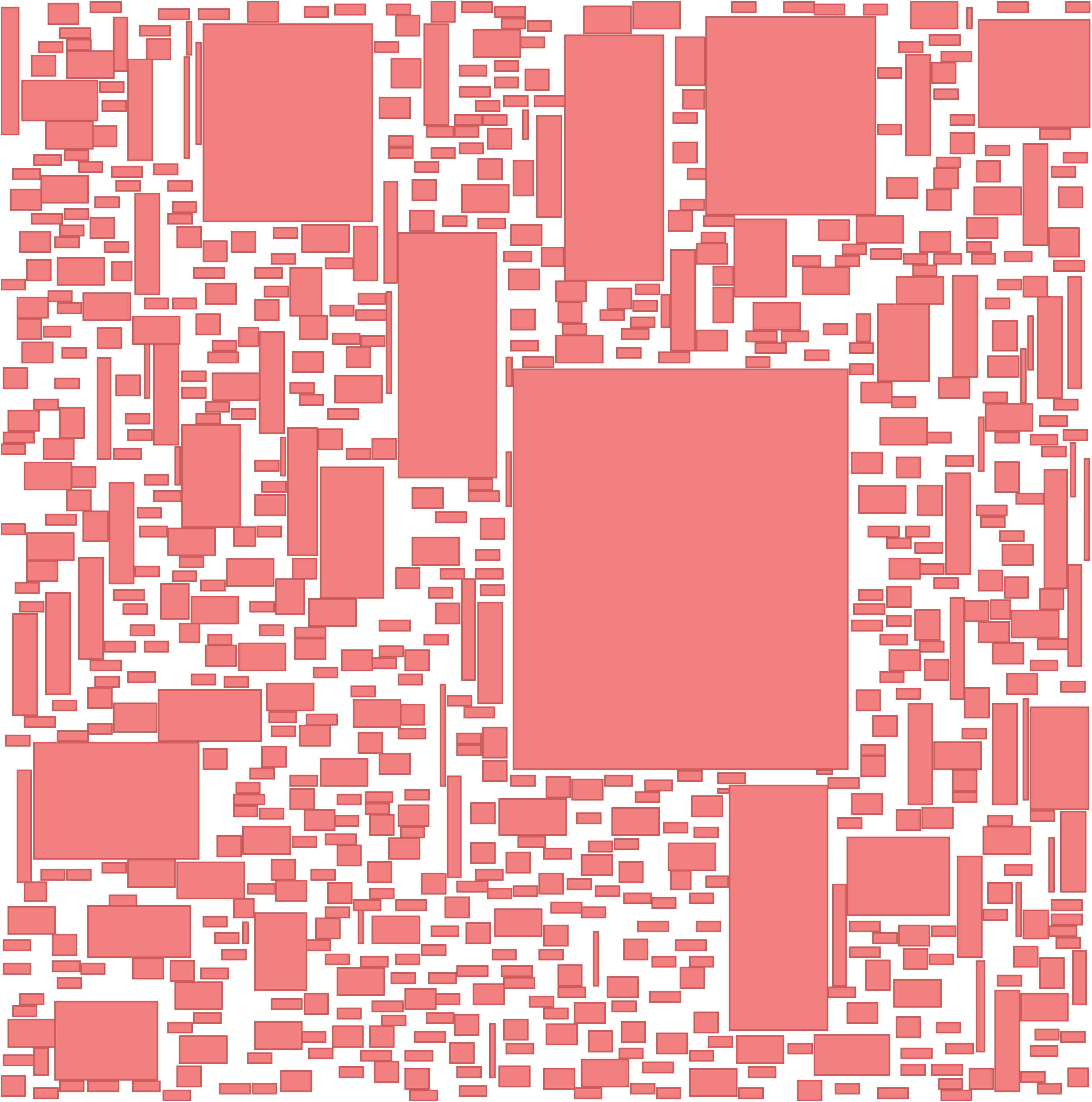}
        } &
        \subfloat[Syn-BigBlue3]{
            \includegraphics[width=0.20\linewidth]{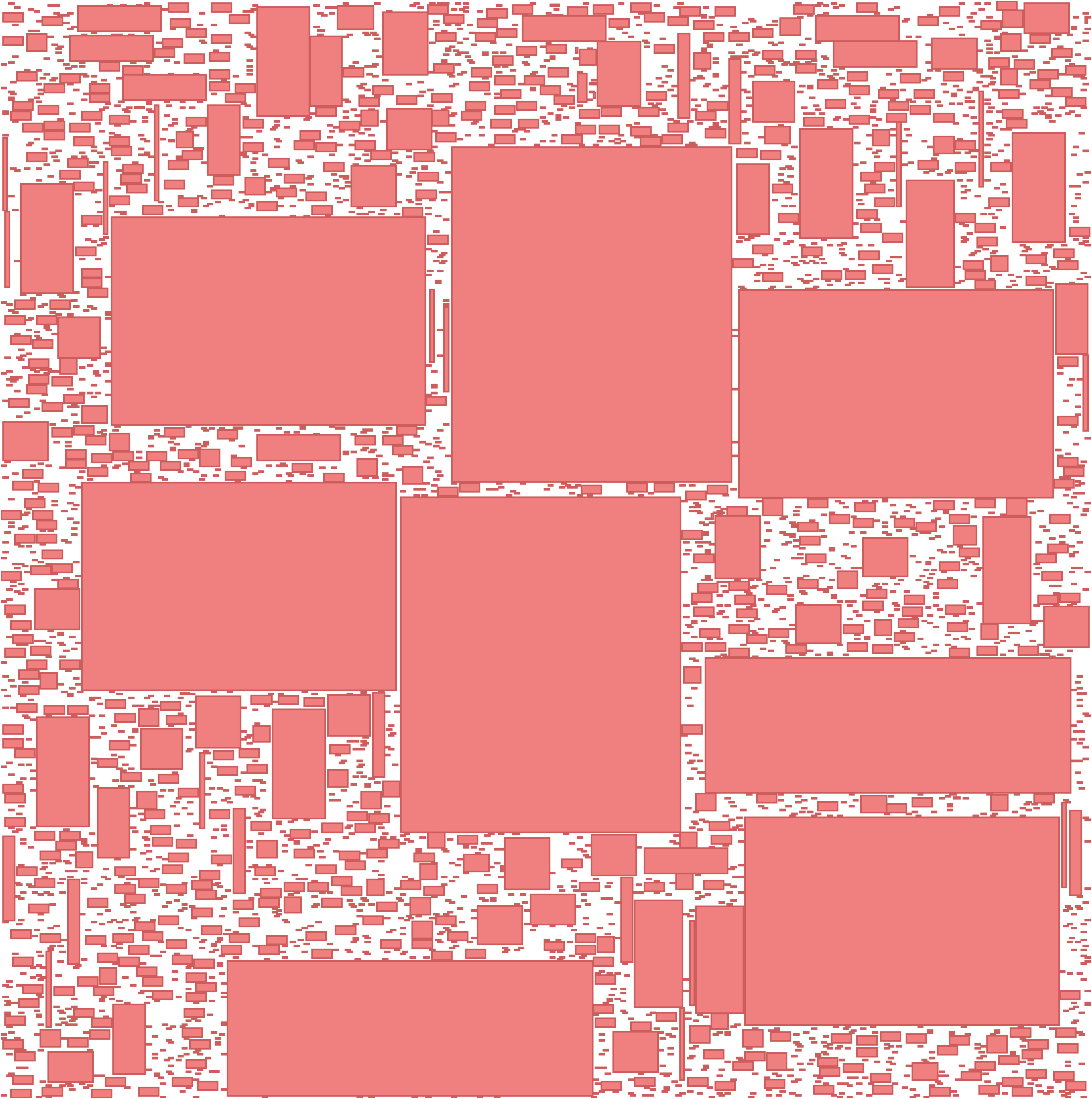}
        } &
        \subfloat[Syn-BigBlue4]{
            \includegraphics[width=0.20\linewidth]{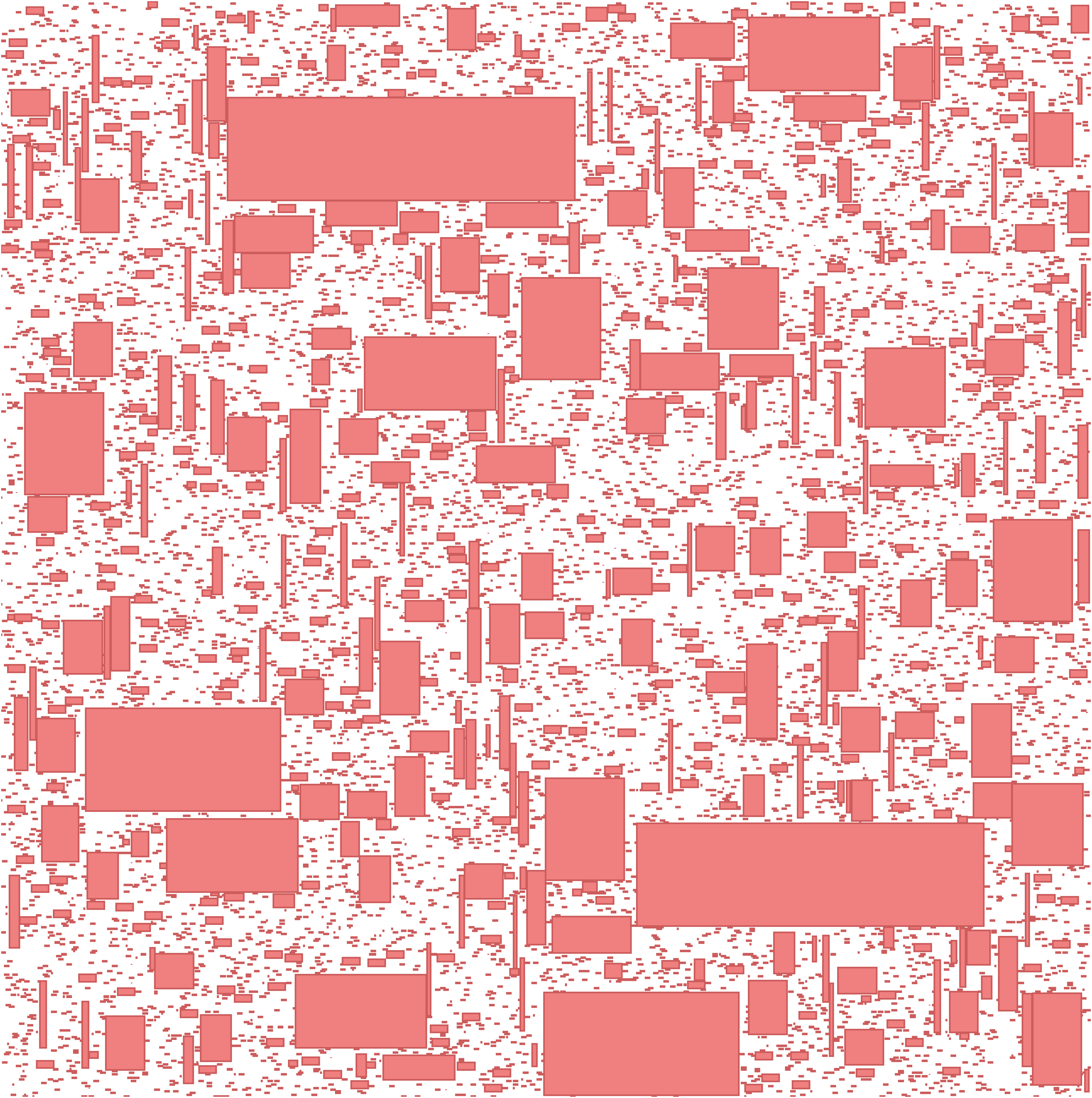}
        } \\[1ex]

        \subfloat[IBM04 (Maskplace)]{
            \includegraphics[width=0.20\linewidth]{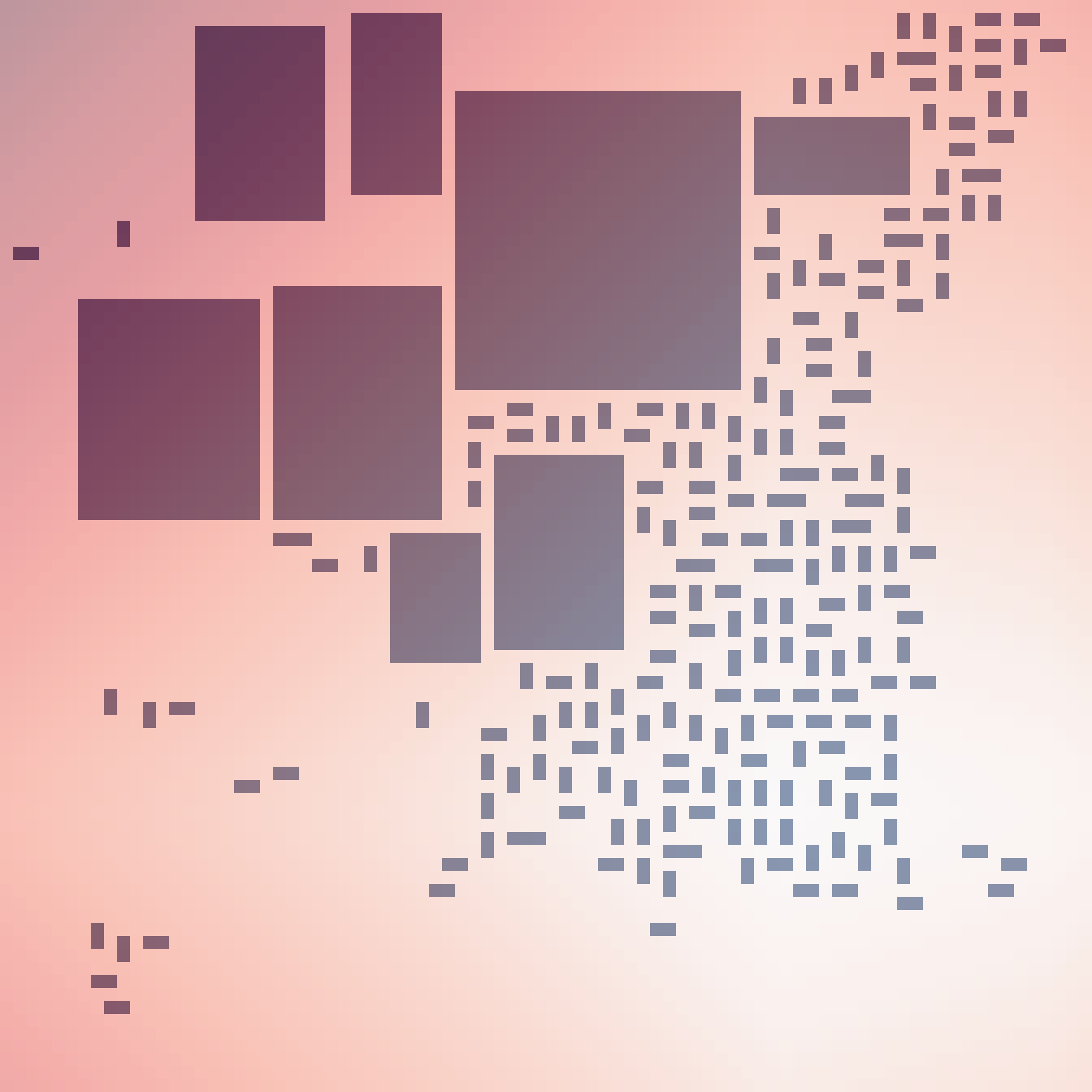}
        } &
        \subfloat[IBM03 (Maskplace)]{
            \includegraphics[width=0.20\linewidth]{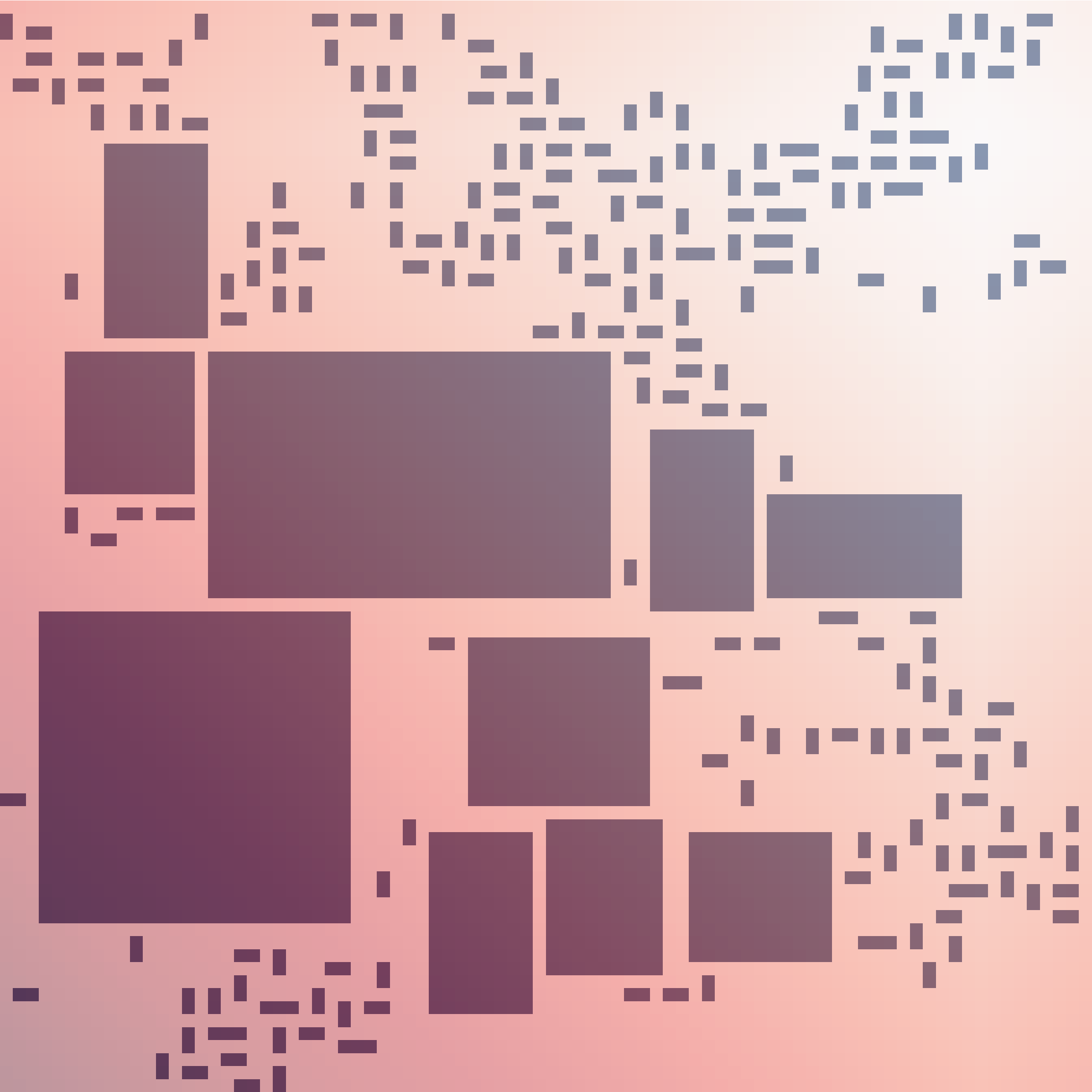}
        } &
        \subfloat[IBM02 (Maskplace)]{
            \includegraphics[width=0.20\linewidth]{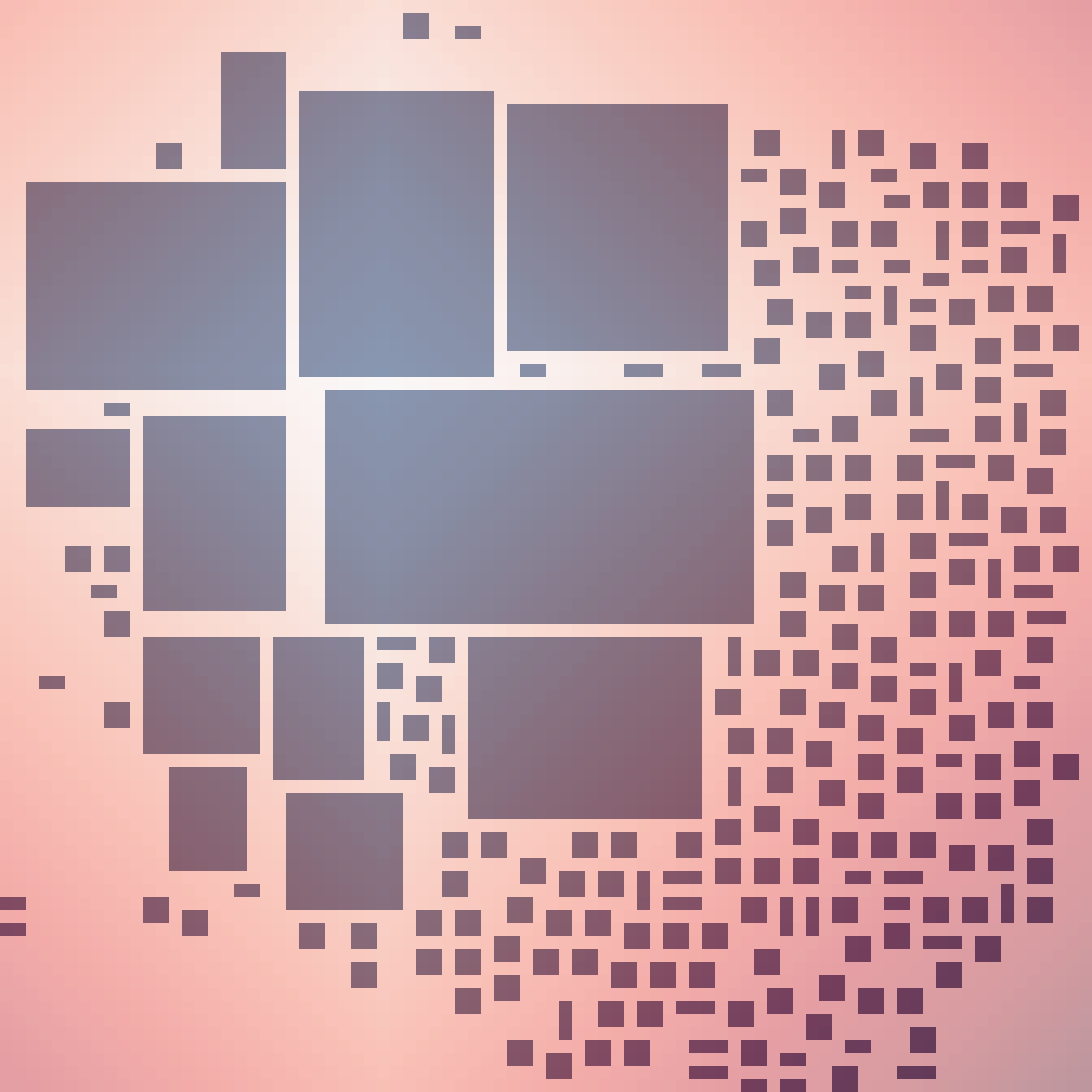}
        }
    \end{tabular}
    \caption{
    Comparison of circuit placement layouts from different data sources.
    Top row: synthetic circuit layouts generated by our proposed data generation pipeline with controllable structural complexity.
    Bottom row: circuit layouts from the public ChipFormer benchmark.
    }
    \label{fig:synthetic-circuit}
\end{figure*}

To ensure robust generalization across diverse circuit topologies, we construct a comprehensive dataset combining real-world benchmarks with generated designs. As illustrated in Figure \ref{fig:synthetic-circuit}, our pipeline augments standard layouts (bottom row) with synthetic instances of controllable complexity (top row), thereby preventing overfitting and expanding the model's exposure to varied macro arrangements.

\subsection{Statistics of Benchmark Circuits}
\label{app:statistics}

Table \ref{tab:benchmark_stats} summarizes the statistics of the benchmark suites used in our evaluation, comprising ISPD05, IBM, and modern TILOS designs. These circuits exhibit a wide range of scale and complexity, with up to 2.2 million standard cells and varying placement densities (11\%--78\%), serving as a robust testbed for assessing scalability and generalization.

\begin{table*}[t]
\caption{Statistics of Benchmark Circuits}
\label{tab:benchmark_stats}
\centering
\footnotesize
\begin{tabular}{lrrrrrrr}
\hline
Circuit & Macros & Hard Macros & Standard Cells & Nets & Pins & Ports & Area Util. (\%) \\
\hline
adaptec1 & 543 & 63 & 210,904 & 221,142 & 944,063 & 0 & 55.62 \\
adaptec2 & 566 & 159 & 254,457 & 266,009 & 1,069,482 & 0 & 74.46 \\
adaptec3 & 723 & 201 & 450,927 & 466,758 & 1,875,039 & 0 & 61.51 \\
adaptec4 & 1,329 & 92 & 494,716 & 515,951 & 1,912,420 & 0 & 48.62 \\
bigblue1 & 560 & 32 & 277,604 & 284,479 & 1,144,691 & 0 & 31.58 \\
bigblue2 & 23,084 & 52 & 534,782 & 577,235 & 2,122,282 & 0 & 32.43 \\
bigblue3 & 1,293 & 138 & 1,095,519 & 1,123,170 & 3,833,218 & 0 & 66.81 \\
bigblue4 & 8,170 & 52 & 2,169,183 & 2,229,886 & 8,900,078 & 0 & 35.68 \\
ariane & 932 & 134 & 0 & 12,404 & 22,802 & 1,231 & 78.39 \\
ibm01 & 256 & 52 & 12,506 & 14,111 & 50,566 & 246 & 61.94 \\
ibm02 & 256 & 52 & 19,321 & 19,584 & 81,199 & 259 & 64.63 \\
ibm03 & 256 & 52 & 22,846 & 27,401 & 93,573 & 283 & 57.97 \\
ibm04 & 256 & 52 & 26,899 & 31,970 & 105,859 & 287 & 54.88 \\
ibm06 & 256 & 52 & 32,320 & 34,826 & 128,182 & 166 & 54.77 \\
ibm07 & 256 & 52 & 45,419 & 48,117 & 175,639 & 287 & 46.03 \\
ibm08 & 256 & 52 & 51,000 & 50,513 & 204,890 & 286 & 47.13 \\
ibm09 & 256 & 52 & 53,142 & 60,902 & 222,088 & 285 & 44.52 \\
ibm10 & 256 & 52 & 68,643 & 75,196 & 297,567 & 744 & 61.40 \\
ibm11 & 256 & 52 & 70,185 & 81,454 & 280,786 & 406 & 41.40 \\
ibm12 & 256 & 52 & 70,425 & 77,240 & 317,760 & 637 & 53.85 \\
ibm13 & 256 & 52 & 83,775 & 99,666 & 357,075 & 490 & 39.43 \\
ibm14 & 256 & 52 & 146,991 & 152,772 & 546,816 & 517 & 22.49 \\
ibm15 & 256 & 52 & 161,177 & 186,608 & 715,823 & 383 & 28.89 \\
ibm16 & 256 & 52 & 183,026 & 190,048 & 778,823 & 504 & 39.46 \\
ibm17 & 256 & 52 & 184,735 & 189,581 & 860,036 & 743 & 19.11 \\
ibm18 & 256 & 52 & 210,328 & 201,920 & 819,697 & 272 & 11.09 \\
mempool-tile & 136 & 36 & 12,584 & 14,230 & 64,520 & 842 & 68.20 \\
nvdla & 476 & 177 & 183,420 & 192,540 & 684,210 & 2,405 & 55.40 \\
\hline
\end{tabular}
\end{table*}

\subsection{Statistical visualization of generated data.}
\label{app:statistics_gendata}

As illustrated in Fig.~\ref{fig:degree_dist}, the node degree distribution of our synthetic data aligns closely with both the statistics of real-world ISPD benchmarks and the theoretical Power Law ($P(k) \propto k^{-1.85}$) typically observed in large-scale VLSI circuits. This confirms that Algorithm 2 successfully captures the complex connectivity patterns essential for robust pre-training, ensuring the learned policies can effectively transfer to unseen industrial designs.

\begin{figure}[t]
    \centering
    \begin{tikzpicture}
        \begin{loglogaxis}[
            xlabel = {Node Degree $k$},
            ylabel = {Probability $P(k)$ (\%)},
            width=8cm,
            height=5.5cm,
            legend style={at={(0.05, 0.05)},anchor=south west, font=\small},
            grid=major,
            ymin=0.1, ymax=100,
            xmin=1, xmax=100
        ]
            \addplot[name path=upper, draw=none, forget plot] 
                table[x=degree, y expr=\thisrow{synthetic}+\thisrow{error}] {data/degree_distribution.txt};
            
            \addplot[name path=lower, draw=none, forget plot] 
                table[x=degree, y expr=\thisrow{synthetic}-\thisrow{error}] {data/degree_distribution.txt};
            
            \addplot[blue!50, opacity=0.6, forget plot] fill between[of=upper and lower];

            
            \addplot[red, thick, mark=square, mark size=2pt] 
                table[x=degree, y=real] {data/degree_distribution.txt};
            \addlegendentry{Real Circuits (ISPD)}
            
            \addplot[blue, thick, mark=o, mark size=2pt] 
                table[x=degree, y=synthetic] {data/degree_distribution.txt};
            \addlegendentry{Ours (Synthetic)}
            
            \addplot[black, dashed, thick, domain=1:50] {100*x^(-1.85)};
            \addlegendentry{Power Law $k^{-1.85}$}
            
        \end{loglogaxis}
    \end{tikzpicture}
    \caption{\textbf{Statistical validation of generated data.} The degree distribution of our synthetic netlists (Blue) closely follows the trajectory of real ISPD circuits (Red) and the theoretical Power Law (Dashed). \textbf{The shaded region represents the error margin (variance) across the synthetic dataset}, confirming that Algorithm 2 captures realistic topological properties essential for effective pre-training.}
    \label{fig:degree_dist}
\end{figure}